\documentclass[apj]{emulateapj}
\usepackage{color}
\usepackage{times}
\usepackage{graphicx}
\usepackage{amssymb,amsmath}

\providecommand{\abs}[1]{\big \vert#1 \big\rvert}

\shorttitle{Overstable librations of first order mean motion resonances in the context of two massive planets}
\shortauthors{Deck et al.}

\begin{document}

\title{Migration of two massive planets into (and out of) first order mean motion resonances}

\author{Katherine M. Deck\altaffilmark{1,3} and Konstantin Batygin\altaffilmark{1}}

\altaffiltext{1}{Department of Geological and Planetary Sciences, California Institute of Technology, Pasadena, CA}
\altaffiltext{3}{Corresponding author: kdeck@caltech.edu}

\begin{abstract}
We consider the dynamical evolution of two planets orbiting in the vicinity of a first order mean motion resonance while simultaneously undergoing eccentricity damping and convergent migration. Following \citet{Sch}, we include a coupling between the dissipative semimajor axis evolution and the damping of the eccentricities. In agreement with past studies, we find that this coupling can lead to overstability of the resonance and that for a certain range of parameters capture into resonance is only temporary. Using a more general model, we show that whether overstable motion can occur depends in a characteristic way on the mass ratio between the two planets as well as their relative eccentricity damping timescales. Moreover, we show that even when escape from resonance does occur, the timescale for escape is long enough such at any given time a pair of planets is more likely to be found in a resonance rather than migrating between them. Thus, we argue that overstability of resonances cannot singlehandedly reconcile convergent migration with the observed lack of {\it Kepler} planet pairs found near resonances. However, it is possible that overstable motion in combination with other effects such as large scale orbital instability could produce the observed period ratio distribution. 

\end{abstract}
\section{INTRODUCTION}

Analyses of the {\it Kepler} data which take into account observational and instrumental biases indicate that Sun-like stars and stars less massive than the Sun commonly host planets \citep{Fressin, Dressing, MortonSwift}. The known planets tend to be smaller than Neptune, orbit with periods less than $\sim 100$ days, and a significant fraction of them reside in multi-planet systems \citep{Batalha,Mullally,Rowe}. Given that these planetary systems may represent the dominant outcome of planet formation, advancing our understanding of their formation and past dynamical evolution is a major goal of exoplanet science.

Although these planets typically have weakly constrained masses, densities, and orbital elements, there are several broad features of the population which can provide clues as to the formation of these systems. One interesting characteristic is that the period ratios of pairs of adjacent planets do not preferentially lie near mean motion resonances. Initially, this was taken as a clue that large-scale migration caused by dissipative interaction with a gaseous protoplanetary disk (e.g. \citealt{KleyNelson,Baruteau}) did not act in a particularly important way, since rudimentary models of convergent migration between planets robustly predict capture into resonance assuming the migration rate is slow enough and the eccentricities are small enough. This apparent contradiction between migration models and the observed data led to the idea of in-situ formation, where these systems ultimately assembled via the same mechanisms that created the terrestrial planets in our Solar System, but acting at significantly smaller orbital distances (e.g. \citealt{Hansen,Hansen2,Chiang}). 

A second feature of these planets complicates this narrative.  By modeling the composition of the subset of planets with measured masses and radii, \citet{Rogers} showed that the majority of planets with radii larger than $\sim 1.6R_\oplus$ require significant gaseous atmospheres.  Processes which produce volatiles after the formation of a planet (such as out-gassing) are thought to proceed at a slower rate than processes which strip a planet of its volatiles (like photo-evaporation) \citep{Lopez,Rogers}.  This suggests that the fraction of volatile rich planets was larger in the past compared with the observed sample, and it also indicates that these planets formed while the gaseous protoplanetary disk was present.

Therefore, in-situ formation no longer provides an immediate explanation for the observed lack of pairs near mean motion resonances by simply obviating the need for large-scale migration. That is, irrespective of where close-in small planets originated, they would have interacted with their gas-rich natal disks. Therefore it is necessary to find a way to explain the lack of observed near commensurabilities in the context of planetary evolution within a gaseous disk, the presence of which seems required to explain the volatile rich nature of a subset of the planets observed.

Both turbulence in the disk and small residual eccentricities can prevent capture into resonance (\citealt{Adams,Rein,Paardekooper,BatyginRes}). Additionally, recent work has suggested that a more complete treatment of convergent migration and eccentricity damping alone can account for the lack of pairs near resonance. Particularly, \citet{Sch} show that in this more complete model, which takes into account how eccentricity damping affects the semimajor axis evolution, the long-term stability of resonances can be compromised (see also \citealt{Meyer}). That is, capture into resonance occurs but it is only temporary, and escape from resonance occurs on timescales comparable to the eccentricity damping time. Since this is small compared with the time spent migrating between resonances (the semimajor axis decay timescale), the expected result is a distribution of period ratios which disfavors resonant values\footnote{These results were derived for the circular restricted three body problem where the inner planet was treated as a test particle.}.

Given the promise of this idea for reconciling planet migration with the fact that most pairs are not near resonance, in this work we extend the theory of eccentricity dependent orbital migration and consider some immediate consequences of the more complete model. In Section \ref{sec:analytic}, we present simple analytic formulae for the stability of the resonant equilibrium. These expressions generalize the results previously obtained for the circular restricted three body problem. Importantly, we show how these criteria depend on the planetary mass ratio and relative eccentricity damping rates between the two planets. This allows us to understand for which parameter values the instability of the resonance can occur and make predictions for which real systems this might have occurred for. We test these analytic results numerically in Section \ref{sec:numeric}. In Section \ref{sec:discussion}, we discuss the implications of our work with regards to whether or not overstable librations of first order resonances can single-handedly account for the observed period ratio distribution of Kepler planets. The derivation of the formulae presented here is given in the Appendix.

 \section{Overstable librations of first order mean motion resonances}\label{sec:analytic}
We consider a system of two planets of mass $m_1$ and $m_2$ orbiting a star of mass $M_\star$ with periods near the $m$:$m+1$ period commensurability. We assume the orbits are nearly circular and nearly coplanar.

In this regime, the Hamiltonian, which governs the conservative dynamics of the planets, is approximately given by
\begin{align}\label{eqn:fullHam}
H & = -\frac{GM_\star m_1}{2 a_1}-\frac{GM_\star m_2}{2 a_2}-\frac{Gm_1m_2}{a_2}\times \nonumber \\
&\bigg[ f_{m+1,27}(\alpha_{\rm{res}})e_1 \cos{[\theta-\varpi_1]} +\nonumber \\
&[f_{m+1, 31}(\alpha_{\rm{res}})-\delta_{m,1}2\alpha_{\rm{res}}]e_2 \cos{[\theta-\varpi_2]}  \bigg]
\end{align}
where $\theta = (m+1)\lambda_2 -m\lambda_1$ and $a_i$, $e_i$, $\lambda_i$ and $\varpi_i$ are the semimajor axes, eccentricities, mean longitudes, and longitudes of periastron of the two planets. The quantities $f_{m+1,27}$ and $f_{m+1,31}$ are functions of Laplace coefficients \citep[p.~539-556]{MurrayDermott} evaluated at $\alpha \equiv a_1/a_2 =\alpha_{\rm{res}}$, where $\alpha_{\rm{res}}$ corresponds to exact commensurability. The term appearing when $m=1$ in the coefficient of the term proportional to $e_2$ is an indirect term in the disturbing function. 

This Hamiltonian can be reduced to one degree of freedom through a series of canonical transformations \citep{Sessin,WisdomCtB,Lemaitre}. Since the derivation exists in the literature, we do not reproduce it here, though a rough sketch is given in the Appendix. After performing the appropriate canonical transformations, the Hamiltonian \eqref{eqn:fullHam}, in the region of phase space close to the resonance, takes the following form:
 \begin{align}\label{Ham1}
H' = -\frac{1}{2}(\Phi'-\Gamma')^2-\sqrt{2\Phi'}\cos{\phi}
 \end{align}
where 
\begin{align}\label{parameters}
\Phi' &\approx \frac{1}{2}\bigg(\frac{3.75m}{\epsilon_p}\bigg)^{2/3}\sigma^2 \nonumber \\
\sigma^2 &\approx e_1^2+e_2^2-2e_1 e_2 \cos{(\varpi_1-\varpi_2)}\nonumber \\
\phi & = (m+1)\lambda_2-\lambda_1 +\psi \nonumber \\
\Gamma' & \approx \frac{1}{2}\bigg(\frac{3.75m}{\epsilon_p}\bigg)^{2/3}\bigg(\sigma^2+\frac{\Delta \alpha}{m}\bigg) \nonumber \\
\tan{\psi} & =- \frac{e_1\sin{\varpi_1}-e_2\sin{\varpi_2}}{e_1\cos{\varpi_1}-e_2\cos{\varpi_2}}
\end{align}
where $\Delta \alpha = \alpha-\alpha_{\rm{res}}$, $\epsilon_p = (m_1+m_2)/M_\star$, $\psi$ is a generalized longitude of pericenter, and $\Gamma'$ is the ``proximity parameter" which governs how close the system is to resonance. All of the above including the Hamiltonian $H'$ are dimensionless. Additionally, the definitions in Equations \eqref{parameters} assume the orbits are compact; in this limit, $\alpha \rightarrow 1,  m\approx m+1$, and $|f_{27}| \approx f_{31} \approx 0.8m$. Full expressions, without this ``compact orbits" approximation, are given in the Appendix.

As has been pointed out before, the functional form of the Hamiltonian in Equation \eqref{Ham1} is identical to that of the circular restricted three body problem (CR3BP) near a first order mean motion resonance, though the parameters have a different meaning. This specific case of the CR3BP has been studied at length and functions as a common analytic model for resonances of various types (see e.g. \citet{HenrardAgain, MurrayDermott,FerrazBook}). It is worth noting that in the ``compact approximation" limit, the mass ratio between the planets $\zeta = m_1/m_2$ does not appear in the Hamiltonian.

\begin{figure}
	\begin{center}
	\includegraphics[width=3.5in]{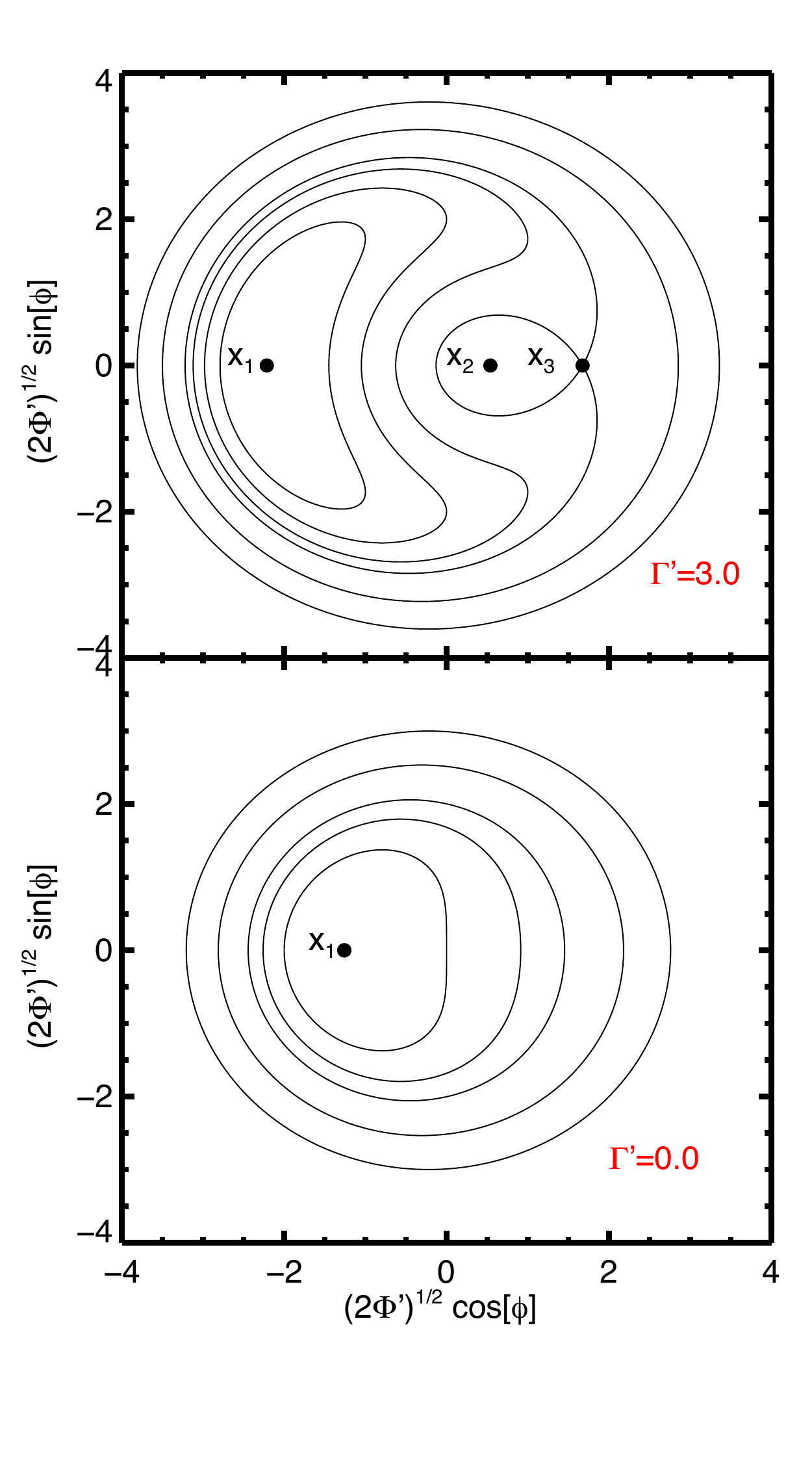}
	\caption{Level curves of the conservative Hamiltonian for two different values of $\Gamma'$. Upper panel: three fixed points as $\Gamma'>3/2$, lower panel: a single fixed point as $\Gamma'\leq 3/2$. }
	\label{fig:contours}
	\end{center}
	\end{figure}

Without dissipation, the proximity parameter is conserved.  When eccentricities are zero ($\sigma=0$), and the orbits are wider than the commensurability ($\alpha <\alpha_{\rm{res}} ), \Gamma'$ is negative, while if the orbits are narrow of the commensurability ($\alpha>\alpha_{\rm{res}} $), $\Gamma'$ is positive. When $\Gamma'<3/2$, the conservative system has a single fixed point ($x_1$), while for $\Gamma'>3/2$ there are three fixed points and a separatrix is present. In this case, two of the fixed points are stable ($x_1$ and $x_2$) and a third ($x_3$) is unstable. To illustrate this, in Figure \ref{fig:contours} we show the level curves of the Hamiltonian given in Equation \eqref{Ham1} for two different values of the proximity parameter $\Gamma'$.  For larger values of $\Gamma'$, the unstable fixed point and the fixed point at the center of resonance correspond to approximately equal values of $\Phi'$ while the second stable fixed point is near zero eccentricity with $\Phi'\approx 0$.  	

In the dissipative case, the equations of motion with respect to the dimensionless time $t'$ can be written symbolically as
\begin{align}\label{EOM}
\frac{d\Phi'}{dt'} & = -\frac{\partial H'}{\partial \phi}+ \frac{d\Phi'}{dt'}\bigg\lvert_{\rm{dis}} \nonumber \\
\frac{d\phi}{dt'} & = \frac{\partial H'}{\partial \Phi'}\nonumber \\
\frac{d\Gamma'}{dt'} & =  \frac{d\Gamma'}{dt'}\bigg\lvert_{\rm{dis}} 
\end{align}
In the physical regime of interest, the dissipation acts on timescales much longer than the natural timescales of the resonance, meaning that the dissipative terms (subscript ``dis") are comparatively small in magnitude. We therefore expect the system to track on short timescales the level curves of the conservative problem with the instantaneous value of $\Gamma'$ and total energy $H$.

The extra terms in the equations of motion \eqref{EOM} are determined by taking the dissipative evolution of the semimajor axes and eccentricities of the planets and converting them into the evolution of $\Phi'$ and $\Gamma'$\footnote{At the order of eccentricity we are working, the angle $\phi$, which depends on eccentricities through the generalized pericenter $\psi$, does not change due to non-conservative effects (see Appendix for proof).}. We parametrize the dissipation that acts on the orbit of each individual planet as
  \begin{align}\label{damping}
\frac{1}{e_i}\frac{de_i}{dt} & = -\frac{1}{\tau_{e,i}}  \nonumber \\
\frac{1}{a_i}\frac{d a_i}{d t} & = \bigg(-\frac{ 2p  e_i^2}{\tau_{e,i}} -\frac{1}{\tau_{a,i}}\bigg),
\end{align}
The above expressions apply for eccentricity decay and migration towards the star ($\tau_{a,i}$ and $\tau_{e,i}$ are positive; to reverse the direction of either the sign of $\tau$ can be changed). These timescales can be estimated analytically (e.g. \citealt{GoldreichTremaine,Tanaka1,Tanaka2}) or from numerical simulations (e.g. \citealt{KleyNelson}) or taken to be free parameters.  Typically $\tau_e\ll\tau_a$ for Type I migration of small planets which do not open a gap in the disk.  In the case of $p \neq 0$, there is a coupling between the semimajor axis evolution and that of the eccentricity evolution, as described by \citet{Sch}. When $p=1$, the eccentricity damping alone exactly conserves the angular momentum of each planet, which then decay slowly on timescales of $\tau_{a,i}$, though estimates indicate $p<1$ \citep{Tanaka2}. In past studies, a simple exponential damping of the eccentricities and semimajor axes was assumed (e.g. \citealt{LeePeale}), without any coupling ($p=0$). %

 In general, two planets undergoing convergent migration will be caught in resonance if the time required to cross the resonance width due to migration is sufficiently long compared with the libration period of the resonance and if the initial eccentricities of the planets are sufficiently low (e.g. \citealt{HenrardAgain,BatyginRes}). Planets undergoing divergent migration cannot be captured into resonance, so throughout this work we focus exclusively on convergent migration ($\tau_{a,1}>\tau_{a,2}$). However, even if the inner planet originally migrates more quickly than the outer planet, convergent migration can arise after the inner planet reaches the inner disk edge, halts and allows the outer planet to catch up. 

\subsection{Dissipative dynamics without a-e coupling}
If there is only convergent migration ($\tau_{e,i} \rightarrow \infty$), the dissipative terms are
\begin{align}\label{eqn:dotgamma1}
\frac{d\Gamma'}{dt}\bigg\lvert_{\rm{dis}}  &=  \frac{a_0}{\tau_{a}} \nonumber \\
\frac{d\Phi'}{dt}\bigg\lvert_{\rm{dis}}  &=  0
\end{align}
where $a_0$ is a positive constant derived in the Appendix (see Equations \eqref{a_defn}, Equations \eqref{EOM_CONSTANTS}) and
\begin{align}\label{derivtau_a}
\frac{1}{\tau_a} & = \frac{1}{\tau_{a,2}}-\frac{1}{\tau_{a,1}}
\end{align} 
For convergent migration,  $\tau_a>0$ and therefore $\Gamma'$ grows as $\alpha \rightarrow \alpha_{\rm{res}}$ (since $\sigma \approx 0$ initially to ensure capture, $\Gamma'$ becomes less negative). Once the system is caught in resonance, $\alpha \approx \alpha_{\rm{res}}$ and the increase in $\Gamma'$ must be compensated with an increase in the eccentricities through an increase in $\sigma$ (Equation \eqref{parameters}).  Without explicit eccentricity damping, the area enclosed by a contour of the Hamiltonian is an adiabatic invariant \citep{Henrard1982}. As $\Gamma'$ grows, the contour which matches the initial area of the trajectory corresponds to larger eccentricities and smaller libration amplitudes. 

When eccentricity damping is included (but $p=0$), the slow evolution of $\Gamma'$ includes an additional damping term, as does that of $\Phi'$, that is,
\begin{align}\label{eqn:dotgamma}
\frac{d\Gamma'}{dt}\bigg\lvert_{\rm{dis}}  \approx  \frac{a_0}{\tau_{a}} +  \frac{c_0}{\tau_{e}}\Phi' \nonumber \\
\frac{d\Phi'}{dt}\bigg\lvert_{\rm{dis}}  \approx \frac{c_0}{\tau_{e}}\Phi'
\end{align}
where $c_0$ is a negative constant derived in the Appendix (see Equations \eqref{c_defn} and Equations \eqref{EOM_CONSTANTS}) which denotes a decay due to eccentricity damping and 
	\begin{align}\label{derivtau_e}
\frac{1}{\tau_{e} }& = \frac{1}{\tau_{e,1}}+\frac{\zeta}{\tau_{e,2}}
\end{align}
where 
\begin{align}\label{zetadefn}
\zeta &= m_1/m_2.
\end{align}
Note first that now the mass ratio between the two planets enters the equations of motion explicitly, while it did not in the conservative case or in the conservative case with only migration. Furthermore, because of the eccentricity damping, there is now a value of $\Phi' = -(a_0/c_0)(\tau_e/\tau_a)$ where the change in $\Gamma'$ is zero.  At a fixed migration rate, a shorter timescale for eccentricity damping coincides with a smaller equilibrium eccentricity since $\Phi'\propto \sigma^2$ . This equilibrium in $\sigma$ corresponds to an equilibrium of the proximity parameter $\Gamma'$ and of $\phi$, and hence the fixed point also corresponds to an equilibrium value of $\alpha$. For weak dissipation, the equilibrium lies close to the stable fixed point in the conservative case (labeled $x_1$ in Figure \ref{fig:contours}), assuming a proximity parameter equal to the equilibrium value of $\Gamma'$ (see Appendix).

The equilibrium in $\sigma$ can be turned into an equilibrium value for $e_1$ and $e_2$ individually by making use of a second conserved quantity $\Psi_2$, which is zero for circular orbits\footnote{Like $\Gamma'$, $\Psi_2$ is conjugate to an angle not appearing in the Hamiltonian, and it is therefore conserved. However, unlike $\Gamma'$, $\Psi_2$ does not appear as a free parameter in the Hamiltonian. See e.g. \citet{ResOverlap} for details.}. The condition that $\Psi_2$ begins at zero (since the orbits begin circular due to damping) and remains zero implies that the pericenters are always anti-aligned, and that the relationship between the equilibrium eccentricities is
\begin{align}
e_2 & = e_1 \zeta \sqrt{\alpha_{\rm{res}}}/R 
\end{align}
where 
\begin{align}\label{Rdefn}
R & \equiv \frac{|f_{m+1,27}(\alpha_{\rm{res}})|}{[f_{m+1,31}(\alpha_{\rm{res}})-2\delta_{m,1}\alpha_{\rm{res}}]}
\end{align}

The stability of the fixed point for the dissipative problem can be determined using linear stability analysis. This yields three eigenvalues (as we have three dimensions, $\Gamma', \Phi'$, and $\phi$), one of which ($\alpha_0$) is always real and always negative, along with a complex conjugate pair of the form $\alpha_\pm = \alpha_1 \pm i \alpha_2$. $\alpha_2$ is well approximated as the libration frequency of the unperturbed problem evaluated at the new fixed point. One finds that $\alpha_1<0$ and so the fixed point is ultimately stable, as has been well established (e.g. \citet{LeePeale}). The action is no longer an adiabatic invariant, however, and the libration amplitude shrinks to zero because the fixed point is a stable attractor.

\subsection{Dissipative dynamics with a-e coupling}\label{sec:overstability}

 \citet{Sch} have shown that the stability of the fixed point is not guaranteed when the semi-major axis evolution is dependent on the eccentricity. The term appearing when $p \neq 0$ changes the dissipative contributions to the equations of motion in Equation \eqref{EOM} at lowest order in eccentricities as:
  \begin{align}\label{eqn:dotgammaB}
\frac{d\Gamma'}{dt}\bigg\lvert_{\rm{dis}}  &\approx  \frac{a_0}{\tau_{a}} +  \bigg[\frac{c_0}{\tau_{e}}+\frac{p a_1}{\tau_{a,e}}\bigg]\Phi' \nonumber \\
\frac{d\Phi'}{dt}\bigg\lvert_{\rm{dis}} & \approx \frac{c_0}{\tau_{e}}\Phi'
\end{align}
where $a_1$ is a negative constant derived in the Appendix (see Equations \eqref{a_defn} and Equations \eqref{EOM_CONSTANTS}), not to be confused with the semimajor axis of the inner planet, and $\tau_{a,e}$ is defined as
\begin{align}\label{tauae}
\frac{1}{ \tau_{a,e}} & = \frac{1}{\tau_{e,1}}-\frac{\zeta^2 \alpha_{res}}{R^2\tau_{e,2}} 
\end{align}
where $R$ is defined in Equation \eqref{Rdefn} and $\zeta$ in Equation \eqref{zetadefn}.

  \subsubsection{Condition for instability}\label{sec:overstable_condition}
Again there is a single fixed point of the dissipative system, as in the case where the coupling parameter $p=0$. The equilibrium value for $\sigma$ of the system, determined by the condition that $d\Gamma'/dt\lvert_{\rm{dis}}=0$ and by the relation between $\Phi'$ and $\sigma$, has a slight shift compared with the case where $p=0$, corresponding to slightly larger or smaller eccentricities depending on if $\tau_{a,e}$ is negative or positive, respectively.  As before, there are three eigenvalues, which we denote as $\alpha_0$ and $\alpha_\pm  = \alpha_1 \pm i \alpha_2$. As in the case when $p=0$, we find that $\alpha_0$ is real and always negative, so that any motion along the associated eigendirection is contracting. 

However, the real part of the complex pair can now be negative or positive. When $\alpha_1>0$, the fixed point is associated with an unstable spiral on the surface spanned by two eigenvectors paired with $\alpha_\pm$. In this case, the lifetime of the system in resonance can be finite. The criterion for $\alpha_1>0$ is given by
 \begin{align}\label{eq:criterion}
\epsilon_p&< \epsilon_{p,crit} \equiv \frac{3pm}{\mathcal{B}  2^{3/2}  }\frac{\tau_{e}}{\tau_{a,e}} \frac{(1+\zeta)^2}{\bigg(m(\zeta+1)+p\frac{\tau_{e}}{\tau_{a,e}}\bigg)^{3/2}} \bigg(\frac{\tau_{e}}{\tau_{a}}\bigg)^{3/2}
 \end{align}
 where $\epsilon_p = (m_1+m_2)/M_\star$, $\mathcal{B} \approx 0.8m$ and $\tau_a$, $\tau_e$, and $\tau_{a,e}$ are as defined in Equations \eqref{derivtau_a}, \eqref{derivtau_e}, and \eqref{tauae}. Please note that this expression employs the compact approximation {\it except} for in $\tau_{a,e}$ where we have retained the factors of $R$ and $\alpha_{\rm{res}}$.

It is interesting that even in the case where the fixed point is associated with an unstable spiral ($\alpha_1>0, \epsilon_p<\epsilon_{p,crit})$, the evolution still drives the system to the fixed point in resonance before the overall instability drives the oscillation amplitude to larger and larger values. Initial capture into resonance requires that the timescale associated with the contracting direction $1/\lvert\alpha_0\lvert$ is much smaller than that of $1/\alpha_1$.  The evolution of the system near the fixed point is made up of a linear combination of the eigenvectors associated with these eigenvalues, and as such, on shorter timescales the contracting evolution dominates. On longer timescales, the growing evolution takes over, and the system can escape from resonance. This separation of timescales is true especially near the critical part of parameter space where the eigenvalue $\alpha_1$ is changing sign, i.e. where the timescale $ 1/|\alpha_1|$ diverges.We will discuss the timescales associated with the evolution further in Section \ref{sec:timescales}.

Regardless of the ultimate stability of the fixed point, the equilibrium eccentricity in the resonance is given by 
\begin{align}\label{eeq}
\sigma_{eq}^2& =  \frac{(1+\zeta)^2}{2m( \zeta+1)+2p\frac{\tau_{e}}{\tau_{a,e}}}\frac{\tau_{e}}{\tau_{a}}. 
\end{align}

As expected, if $p=0$ and there is no coupling between eccentricity damping and semimajor axis evolution, the criterion given in Equation \eqref{eq:criterion} can never be satisfied - the equilibrium in resonance is always stable. The equilibrium eccentricity is still given by Equation \eqref{eeq}.

A sufficient criterion for stability is $\epsilon_p>\epsilon_{p,crit}$. We assume that convergent migration leads to resonance capture and that eccentricity damping leads to an equilibrium in resonance.  Therefore the parameters $\tau_{e}$ and $\tau_{a}$ are positive. Note that for the equilibrium eccentricity to be real - for the equilibrium to even exist -  the denominator in Equation \eqref{eeq} must be positive, which implies that the factor $ (m(\zeta+1)+p\tau_{e}/\tau_{a,e})^{3/2} $ appearing in the critical value $\epsilon_{p,crit}$ for overstable librations is also real and positive. 

This implies that the criterion cannot be satisfied if $\tau_{a,e}<0$, since in that case $\epsilon_{p,crit}$ is negative. This occurs when
 \begin{align}\label{tau_ae}
\frac{\tau_{e,2}}{\tau_{e,1}}<\zeta^2 \frac{\alpha_{\rm{res}}}{R^2}
 \end{align}

 	\begin{figure}
	\begin{center}
	\includegraphics[width=2.5in]{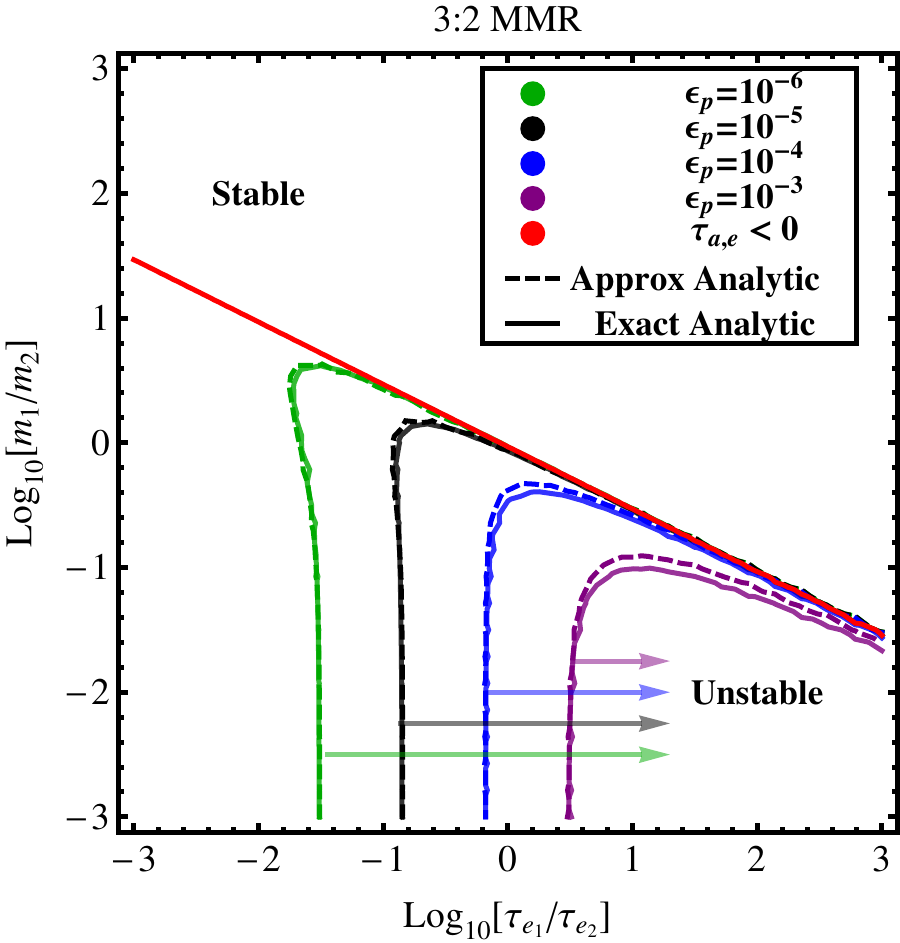}
	\caption{Critical curves showing where the fixed point of the 3:2 resonance is associated with an unstable spiral as a function of $\tau_{e,1}/\tau_{e,2}$ and $\zeta = m_1/m_2$. The different curves green, black, blue, and purple correspond to varying $\epsilon_p$, while the red curve shows where $\tau_{a,e}=0$. The region above the red curve has $\tau_{a,e}<0$ and is stable. Below this curve, the fixed point can be unstable, but only if the parameters lie in the lower right region, below the critical curves. The dashed lines show the predictions after making the compact approximation (though not for $\tau_{a,e}$).  The eccentricity damping timescale of the outer planet is fixed at 100x shorter than the migration time $\tau_{a,2}$, and $\tau_{a,1}\sim 500\tau_{a,2}$. }
	\label{fig:32}
	\end{center}
	\end{figure}
		\begin{figure}
	\begin{center}
	\includegraphics[width=2.5in]{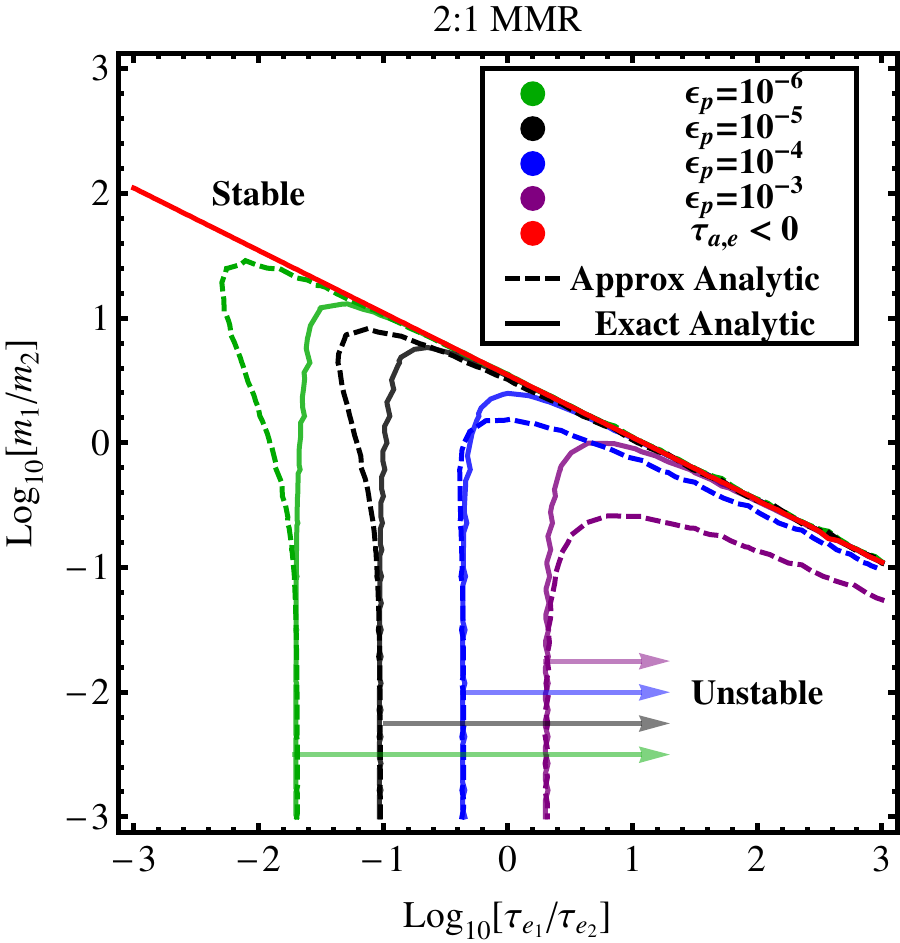}
	\caption{Critical curves showing where the fixed point of the 2:1 resonance is associated with an unstable spiral as a function of $\tau_{e,1}/\tau_{e,2}$ and $\zeta = m_1/m_2$. Refer to Figure \ref{fig:32} for details. }
	\label{fig:21}
	\end{center}
	\end{figure}
 In Figure \ref{fig:32} and Figure \ref{fig:21}, we show the critical curves governing the stability of the fixed point for the 3:2 and the 2:1 resonances, defined by Equation \eqref{eq:criterion}, on the parameter plane of $\tau_{e,1}/\tau_{e,2}$ and $\zeta = m_1/m_2$. In these plots, we have chosen $\tau_a \approx \tau_{a,2}$  $(\tau_{a,1} \rightarrow \infty)$ and $\tau_{e,2} = \tau_{a,2}/100$. There are no other free parameters.
  
Each plot shows the following.  The red curve denotes $\tau_{a,e} = 0$, and the area above the red curve corresponds to a region of parameter space where the fixed point is stable since $\tau_{a,e,}<0$. Below the red curve, the fixed point may be unstable, but only if $\epsilon_p<\epsilon_{p,crit}$. The green, black, blue, and purple curves correspond to $\epsilon_p=\epsilon_{p,crit}$ for $\epsilon_p =  (10^{-3},10^{-4},10^{-5},10^{-6})$, respectively, and the area to the lower right of these curves is where the instability occurs. 
 
 In order to simplify the expressions for the critical values of e.g. $\epsilon_p$ to that given in Equation \eqref{eq:criterion}, we made the ``compact approximation" that $\alpha =a_1/a_2 \rightarrow 1$ as discussed above (note again that we do not make that approximation for $\tau_{a,e}$). This approximation is poorest for the 2:1 resonance, both because $\alpha_{\rm{res}}$ is further from unity but also because of the indirect contribution to the coefficient of the $e_2$ term in the disturbing function, so that $R=2.78$.  However for closer resonances the approximation is very good. 
 
 The difference between the dashed (approximate) and solid (exact) $\epsilon_p=\epsilon_{p,crit}$ curves in the Figures demonstrates this.  Note that the disagreement would be stronger for the 2:1 resonance if we had used the compact approximation for $\tau_{a,e}$, which is why we keep the full expression (it is easy to include it here and retain a simple expression for $\epsilon_{p,crit}$; the same is not true if we had never made the approximation at all). Regardless, the exact formulae is reasonably well approximated by the estimate, and both show the basic result that systems with a more massive planet interior are more stable against overstable librations. This is especially striking when we consider the two cases of an inner and an outer test particle below.

 \subsubsection{Limiting case of CR3BP}
Given that these results were derived within the framework of the elliptic planetary three body problem,  they should reduce to the results obtained previously in the limit of the circular restricted three body problem. We first consider the case of a test particle moving outwards towards a massive planet.  In this case, $\zeta \rightarrow 0, \tau_{a,2} \rightarrow \infty, \tau_{e,2}\rightarrow \infty$, and $\tau_{a,1} \rightarrow -\tau_{a,1}$ (to account for outward migration). Then $\tau_{e} = \tau_{a,e} = \tau_{e,1}$ and $\tau_{a} = -\tau_{a,1}$. We also assume the outer planet has a circular orbit, so $\sigma = e_1$. In this case, when $p=1$ the equilibrium eccentricity is given by 
\begin{align}
e_1^2& =  \frac{1}{2(m+1)}\frac{\tau_{e,1}}{\tau_{a,1}} = \frac{1}{3(m+1)}\frac{\tau_{e,1}}{\tau_{n,1}} \nonumber \\
 \end{align}
 where $\tau_{n,1} =2/3\tau_{a,1}$. This agrees with Equation (24) of \citet{Sch}.
 The criterion for overstability is
  \begin{align}\label{eq:criterionTP_Kat}
\epsilon_2&< \frac{m}{\mathcal{B}   } \frac{}{\sqrt{3}(m+1)^{3/2}} \bigg(\frac{\tau_{e,1}}{\tau_{n,1}}\bigg)^{3/2}
 \end{align}
 which agrees with Equation (30) of \citet{Sch}\footnote{In their formulation, they use $\tau_{n}$ rather than $\tau_{a}$. Additionally, though we have used the same symbol for the coupling parameter $p$, our case with $p=1$ corresponds to their case with $p=3$. Finally, we are using a capital $\mathcal{B}$ to represent $0.8m$, which they use a lowercase $\beta$ for, because we use a lowercase $\beta$ for a different meaning in the derivation in the Appendix.}.
 
 We now turn to the opposite case of a test particle moving inwards towards a massive planet.  In this case, $\zeta \rightarrow \infty, \tau_{a,1} \rightarrow \infty$ and $ \tau_{e,1}\rightarrow \infty$. Then $\tau_{e} = \tau_{e,2}/\zeta,  \tau_{a} = \tau_{a,2}$, and $\tau_{a,e} = -\tau_{e,2}/\zeta^2$. The fact that  $\tau_{a,e}<0$ in this case immediately implies that the resonance is stable. Indeed, after carefully taking the limit as $\zeta \rightarrow \infty$, the criterion for over stability is
  \begin{align}\label{eq:criterionTPouter}
\epsilon_{1} <  -\frac{3pm}{\mathcal{B}  2^{3/2}  } \frac{1}{(m-p)^{3/2}} \bigg(\frac{\tau_{e,2}}{\tau_{a,2}}\bigg)^{3/2}
 \end{align}
 which can never be satisfied. When $m\leq p$, this expression either diverges or becomes imaginary. However, in these cases, the equilibrium itself does not exist (see Appendix). Note also that the divergence when $m=p=1$ does not occur within the full expression (i.e. without taking the compact limit).

Why is there a difference between the two limiting cases? We do not yet have a good physical intuition for this. However, since the Hamiltonian itself is approximately independent of the mass ratio between the two planets, any dependence on the mass ratio must come from the dissipative terms. In the case where the inner planet becomes a test particle, the eccentricity damping leads to an inward migration of the test particle proportional to $e_1^2$, in opposition to the overall outward migration towards the massive planet. When the outer planet is the test particle, the eccentricity dependent migration acts coherently with the direct semimajor axis damping to move the test particle towards the inner planet. The small contribution of $p$ in one case apparently compromises the stability of the resonance, while in the other it stabilizes it further. 
  \subsubsection{Condition for escape from resonance}\label{sec:limitcycle}

If the fixed point is an unstable spiral, the libration amplitude about the fixed point will grow in time. For a range of parameter values, these oscillations will saturate at a stable limit cycle enclosing the unstable fixed point. In this case, the system remains trapped in resonance but with a nonzero libration amplitude. A criterion for escape from the resonance would be such that the fixed point is unstable and there is no possibility of saturating at a stable limit cycle.

We can understand this qualitatively as follows. The true fixed point of the dissipative problem lies near the fixed point of the conservative problem at the center of the resonance region (labeled $x_1$ in Figure \ref{fig:contours}). There exists a limit cycle because the motion is being driven towards the fixed point in one eigendirection and away from the fixed point as an unstable spiral in the other two eigendirections, and there is a balance of these opposing actions at some point.   However, when there are three fixed points of the conservative problem ($\Gamma'>3/2$), and the fixed point of the full dissipative problem is an unstable spiral, the libration amplitude will grow until the trajectory enters a region of attraction in the $(\sqrt{2\Phi'}\cos{\phi},\sqrt{2\Phi'}\sin{\phi})$ plane near the conservative fixed point $x_2$ at $\sigma\approx0$, without reaching a stable limit cycle. 
 
 This region of attraction corresponds to what would have been the inner circulation region (corresponding to oscillations about $x_2$).  The point $x_2$ is {\it not} a fixed point of the full problem, but the motion in the $(\sqrt{2\Phi'}\cos{\phi},\sqrt{2\Phi'}\sin{\phi})$ plane near this region appears as a spiral towards $x_2$,  because the dissipative evolution approximately follows contours of the conservative problem on short timescales. For large values of $\Gamma'$, the stable fixed point ($x_2$) of the conservative problem corresponds to nearly zero eccentricity $\sigma$, and so in this attractive region the eccentricities of the planets damp to nearly zero eccentricity.   $\Gamma'$ continues to grow, since this is not a fixed point of the dissipative problem, and this brings the eccentricities closer to zero and brings the pair narrow of the resonance (since for $\sigma \sim 0$ a positive $\Gamma'$ implies $\alpha>\alpha_{\rm{res}} $, see Equation \eqref{parameters}).  This evolution is illustrated in Section \ref{num:illustrate} where we show the numerically determined evolution of $\Phi'$ and $\phi$ on the $(\sqrt{2\Phi'}\cos{\phi},\sqrt{2\Phi'}\sin{\phi})$ plane along with appropriate contours of the conservative Hamiltonian.

 Applying this  criterion, we find that the system avoids being trapped in a limit cycle if $\epsilon_p<\epsilon_{p,crit}$ {\it and} $\Gamma' >3/2$ or 
 \begin{align}\label{eqn:LC}
 \epsilon_p&\lesssim \frac{3m^2}{16\mathcal{B}  \sqrt{2}  } \frac{(1+\zeta)^3}{\bigg(m(\zeta+1)+p\frac{\tau_{e}}{\tau_{a,e}}\bigg)^{3/2}} \bigg(\frac{\tau_{e}}{\tau_{a}}\bigg)^{3/2}\equiv \epsilon_{LC} \nonumber \\
 \bigg(\epsilon_{LC} & = \frac{m}{8}\frac{\tau_{a,e}}{\tau_e}\frac{(1+\zeta)}{p}\epsilon_{p,crit}\bigg)
 \end{align}
 
  If 1) $\epsilon_{p,crit} < \epsilon_{p}$ or  2) $\epsilon_{LC} < \epsilon_p < \epsilon_{p,crit}$ the system is stuck in resonance, either at the (stable) fixed point in the former case or in a limit cycle about the (unstable) fixed point in the latter. Note that the critical value $\epsilon_{LC}$ is nonzero even if the coupling parameter $p=0$. This doesn't mean that the system can escape from resonance if $\epsilon_p<\epsilon_{LC}$ even if $p=0$, but that this limit cycle criterion is meaningless unless $\epsilon_p<\epsilon_{p,crit}$ in the first place. Finally, as pointed out in \citet{Sch}, the limit cycle is only a factor for resonances satisfying $\epsilon_{LC}<\epsilon_{p,crit}$. For an inner test particle with $p=1$, this corresponds to $m<8$.  Because of this, we focus on Equation \eqref{eq:criterion} as a criterion for instability, though formally one requires $\epsilon_p <\epsilon_{LC} <\epsilon_{p,crit}$ for escape from resonance.

 \section{Numerical Results}\label{sec:numeric}
 Here we test how well our simple analytic criterion applies to a ``real" system using direct numerical integration of the full gravitational equations of motion with the appropriate dissipative terms put in.  We integrate the standard gravitational equations of motion using a Bulirsch-Stoer integration scheme. The migration terms are added directly to the equations of motion following the prescription in the Appendix of \cite{LeePeale}. This requires applying the chain rule to determine how changes in $a$ and $e$, defined in Equation \eqref{damping}, translate into changes in the cartesian positions and velocities. As the planets migrate towards the host star, their orbital periods decrease, and an adequate time step for the initial orbits may be too large for the orbits at a later time in the integration. To alleviate this issue and keep the (fixed) dissipation timescales slow compared to the orbital periods, we rescale the semimajor axes of the planets at each time step so that the semimajor axis of the inner planet is fixed. Then our default time step is always short compared to the orbital periods, and the fixed migration rates are typically long compared to the the relevant libration timescales.

\subsection{Illustration of instability effect}\label{num:illustrate}
We begin by showing the explicit evolution of $\Phi'$, $\phi$, and $\Gamma'$ in the three cases of permanent capture with no limit cycle, permanent capture with a limit cycle, and escape from resonance to better illustrate the above discussion of Section \ref{sec:limitcycle}.

\begin{figure}
	\begin{center}
	\includegraphics[width=3.5in]{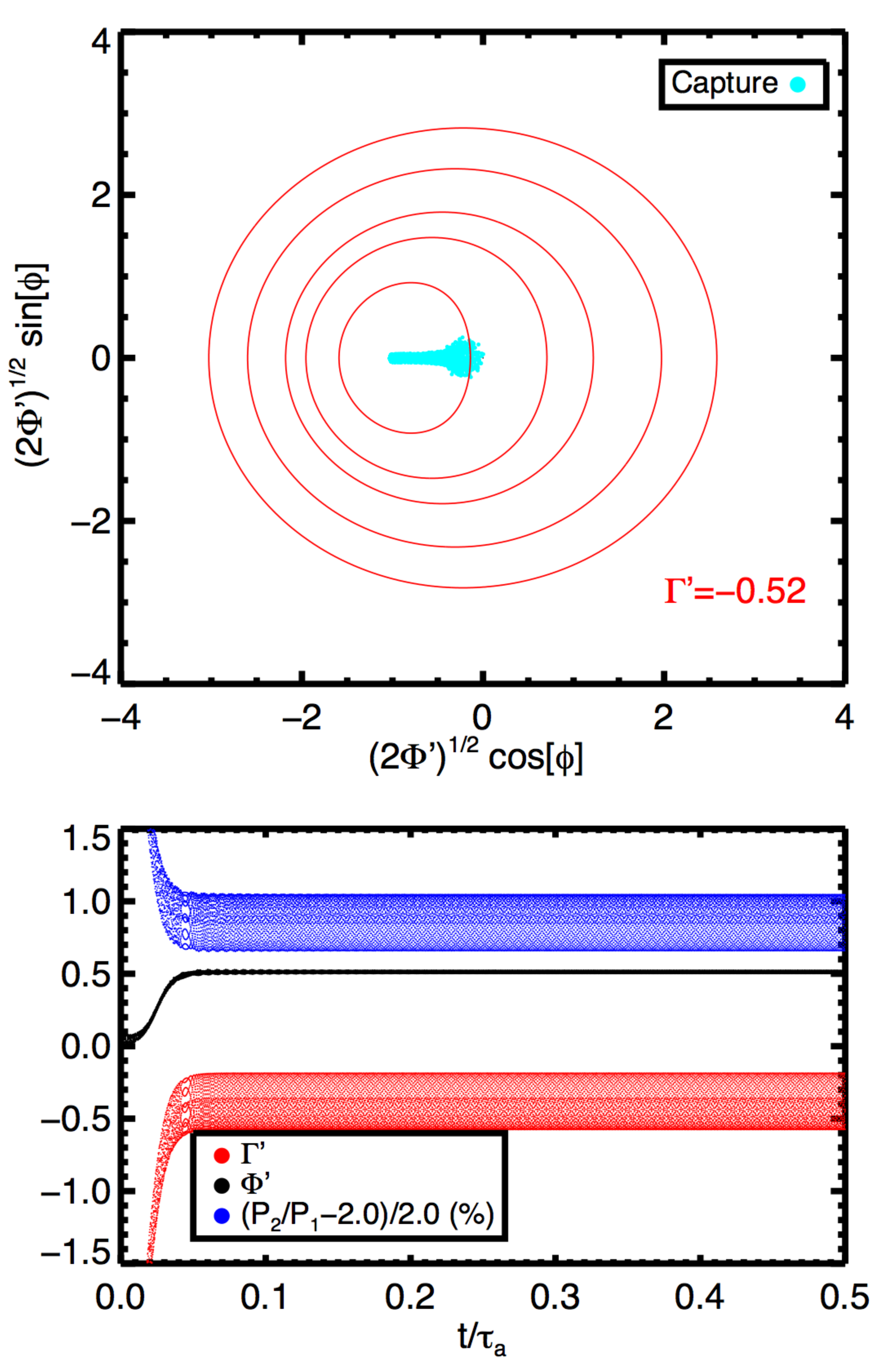}
	\caption{Permanent capture into the 2:1 resonance in the case where the fixed point is stable. Upper panel: contours of the conservative Hamiltonian at the equilibrium value of $\Gamma' \approx -0.52$ (red) and the actual dissipative evolution of the system showing capture into resonance (cyan). Lower panel: time evolution of $\Gamma'$ (red), $\Phi'$ (black), and the fractional deviation of the period ratio from 2.0 in percent (blue). One can see a small damping of the oscillation amplitude of $\Phi'$. }
	\label{fig:capture}
	\end{center}
	\end{figure}

\begin{figure}
	\begin{center}
	\includegraphics[width=3.5in]{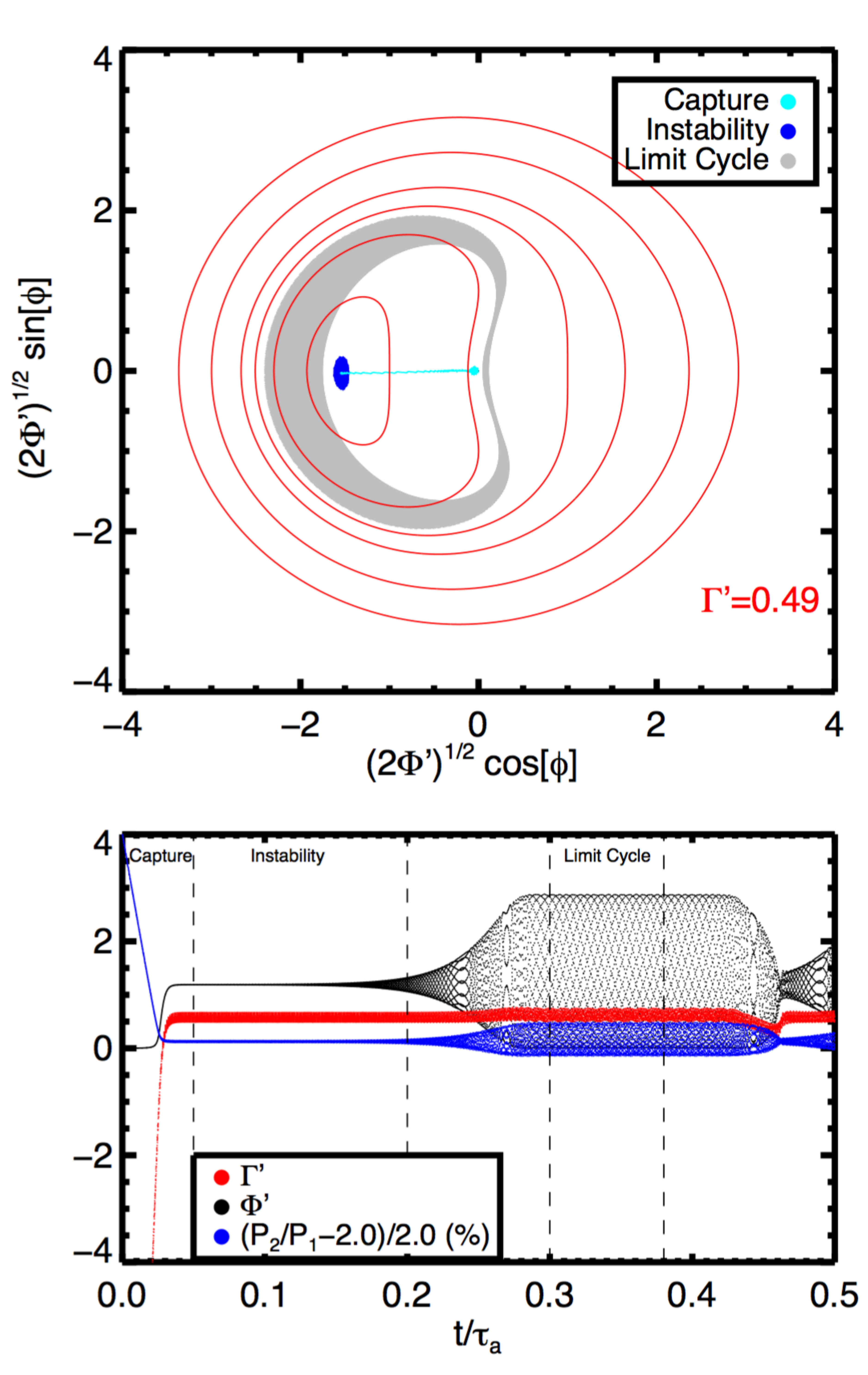}
	\caption{Permanent capture into the 2:1 resonance in the case where the fixed point is unstable. Upper panel: contours of the conservative Hamiltonian at the equilibrium value of $\Gamma' \approx 0.49$ (red) and the actual dissipative evolution of the system showing capture into resonance (cyan) and subsequent growth of the libration amplitude (blue) to a stable limit cycle (grey). We do not show the entire evolution from initial onset of instability to the limit cycle. Lower panel: time evolution of $\Gamma'$ (red), $\Phi'$ (black), and the fractional deviation of the period ratio from 2.0 in percent (blue).  The dashed lines and the regions in between reflect those used for the ``capture", ``onset of instability", and ``escape" evolution in the upper panel. The sudden change in amplitude of oscillations at $\sim0.4\tau_a$ is not captured by our simple analytic formulation.}
	\label{fig:LC}
	\end{center}
	\end{figure}

\begin{figure}
	\begin{center}
	\includegraphics[width=3.5in]{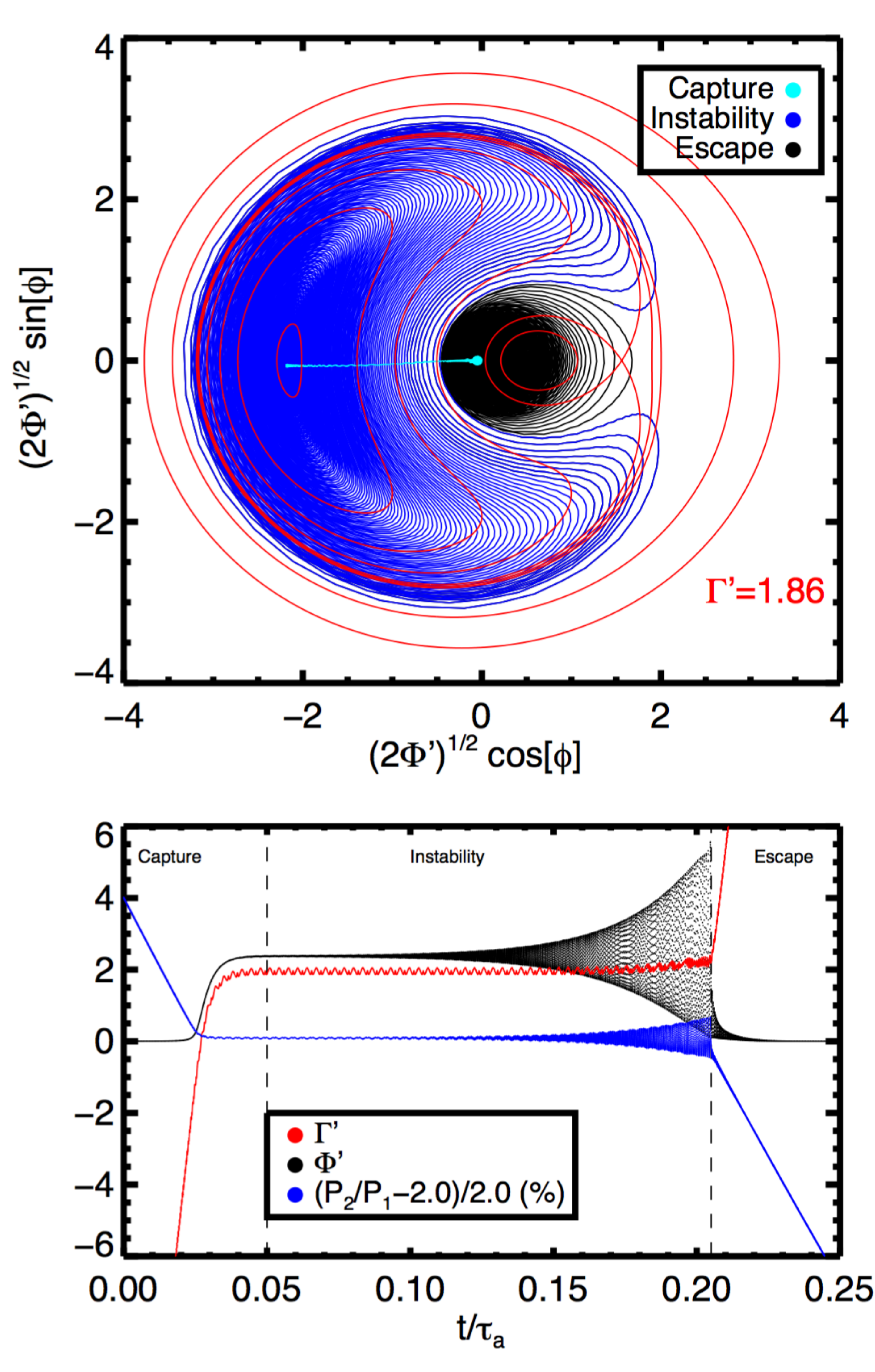}
	\caption{Temporary capture into the 2:1 resonance in the case where the fixed point is unstable. Upper panel: contours of the conservative Hamiltonian at the equilibrium value of $\Gamma' \approx 1.86$ (red) and the actual dissipative evolution of the system showing capture into resonance (cyan), subsequent growth of the libration amplitude (blue), and finally escape from the resonance by damping to second stable fixed point of the {\it conservative} problem (black). See text for an explanation. Lower panel: time evolution of $\Gamma'$ (red), $\Phi'$ (black), and the fractional deviation of the period ratio from 2.0 in percent (blue). The dashed lines and the regions in between reflect those used for the ``capture", ``instability", and ``escape" evolution in the upper panel.}
	\label{fig:escape}
	\end{center}
	\end{figure}

In Figure \ref{fig:capture}, Figure \ref{fig:LC}, and Figure \ref{fig:escape}, we show the evolution of three different systems. We have set $\tau_{e,1} = \tau_{e,2}$, $\zeta = 0.6$, $\tau_{e,2} = 10^4 P_1$, and $\tau_{a,1}\gg \tau_{a,2}$. In case 1 $\epsilon_p = 8\times10^{-4}$ and $\tau_{a,2} \sim 10^6 P_1$, in case 2 $\epsilon_p = 8\times10^{-5}$ and $\tau_{a,2} \sim 2\times 10^6 P_1$, and in case 3 $\epsilon_p = 8\times10^{-5}$ and $\tau_{a,2} \sim 10^6 P_1$. In case 1 and case 3, $\epsilon_{p,crit} = 3\times10^{-4}$ and $\epsilon_{LC}  = 10^{-4}$, while in case 2 $\epsilon_{p,crit} = 10^{-4}$ and $\epsilon_{LC}  = 3.5 \times10^{-5}$. To calculate these values, we used Equations \eqref{eq:criterion} and \eqref{eqn:LC}.

In all cases, the planetary system is captured into the 2:1 mean motion resonance and initially driven to the fixed point, as discussed in Section \ref{sec:overstable_condition}. We show in the upper panel of each of these figures the contours of the conservative Hamiltonian on the  $(\sqrt{2\Phi'}\cos{\phi},\sqrt{2\Phi'}\sin{\phi})$ plane (in red) corresponding to approximate equilibrium value of $\Gamma'$. Overplotted on these contours is the behavior of the system variables undergoing full dissipative evolution. In all three cases, the system begins at zero eccentricity (the origin). The planets are captured into the mean motion resonance and this initially leads to an increase in eccentricities (radial distance from the origin) until the equilibrium is reached (this stage of the evolution is shown with cyan points). 

First, in Figure \ref{fig:capture}, we show case 1, where we have chosen $\epsilon_p>\epsilon_{p,crit}$ so that the fixed point is stable ($\alpha_1<0$). The system remains at the fixed point. In the bottom plot, we show the evolution of $\Gamma'$, $\Phi'$, and the fractional deviation in the period ratio from the exact commensurability. It is difficult to see by eye, but the amplitude of oscillation of $\Phi'$ is decreasing as we would expect.

In Figure \ref{fig:LC}, we show case 2, where we have chosen $\epsilon_{LC}<\epsilon_p<\epsilon_{p,crit}$. The limit cycle behavior is possible because the equilibrium $\Gamma'$ is less than $3/2$, and so there is only one fixed point of the conservative problem.  In the upper panel we now show in dark blue the stage of the evolution where the amplitude of oscillations grows about the fixed point (instability). In grey the limit cycle is shown. Note that it encloses the origin, and therefore the resonant angles are circulating in this configuration. The amplitude of the limit cycle is changing because the proximity parameter $\Gamma'$ is oscillating as well, which changes the level curves of the conservative Hamiltonian.  It is unclear what exactly causes the behavior at a time of $0.4\tau_a$, when the amplitude of oscillations suddenly decreases and then begins to increase again. We note that a precise analytic description of the limit cycle is quite complicated; the limit cycle criterion we use is a heuristic one. 
	
Finally, in Figure \ref{fig:escape}, we show how a system can escape from resonance. In this case, $\epsilon_{p}<\epsilon_{LC}<\epsilon_{p,crit}$ because $\Gamma'$ is greater than $3/2$ in the equilibrium configuration. The amplitude of oscillation grows about the fixed point (shown again in blue) until it crosses into the basin of attraction dominated by the second stable fixed point of the conservative problem. This stage of the evolution is shown in black. At this point, $\Phi'$ begins to decrease while $\Gamma'$ continues to increase (lower panel). As $\Gamma'$ increases, the second stable fixed point of the conservative problem moves closer to the origin and the system eccentricity decreases to zero. This then causes the period ratio to shift to be narrow of the resonance since $\Gamma'$ relates the eccentricity $\sigma$ and the period ratio.

\subsection{Long-term evolution}
		\begin{figure}
	\begin{center}
	\includegraphics[width=3.5in]{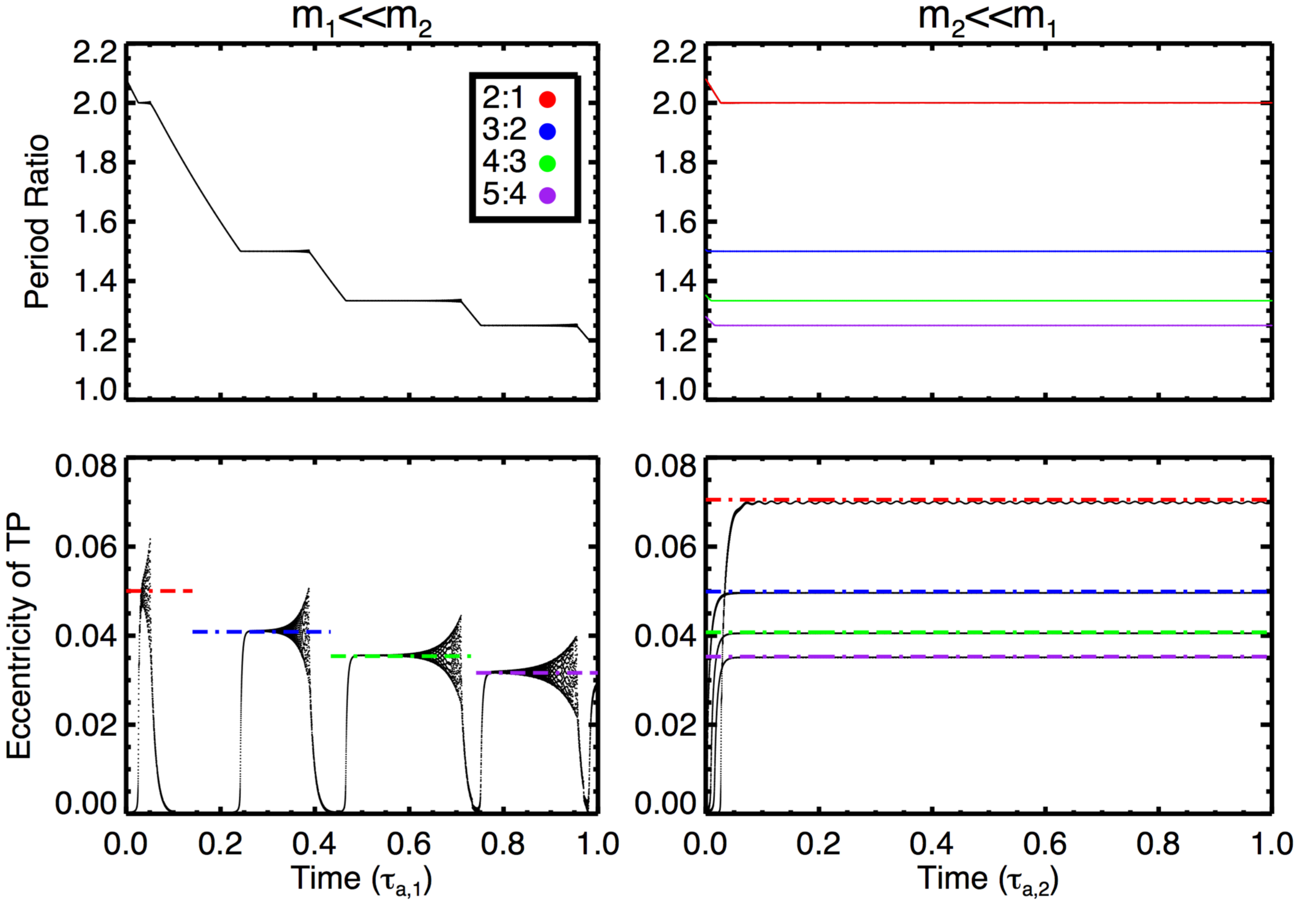}
	\caption{The asymmetry between an outer test particle and an inner test particle. On the left we show the evolution of a single trajectory in period ratio (top) and eccentricity (bottom) of a test particle migrating outwards towards a massive planet on a fixed circular orbit. On the right we show the evolution of four separate orbits, began just outside the 2:1 (red), 3:2 (blue), 4:3 (green), and 5:4 (purple) resonances, for an outer test particle migrating inwards towards a massive planet on a fixed circular orbit.  The dashed colored lines in the bottom two panels show the estimated equilibrium eccentricity of each resonance from the exact expression derived in the Appendix. See text for further details and discussion.}
	\label{fig:testPart}
	\end{center}
	\end{figure}

In Figure \ref{fig:testPart}, we show the evolution of the period ratio and eccentricity $\sigma$ of two ``restricted-like" systems. On the left side, the two plots show the results when the inner planet is effectively a test particle with a mass $10^{-8}$ the mass of the star, while the outer planet is $10^{-5}$ the mass of the star. We choose the migration rate and eccentricity damping of the outer planet to be very long compared with all other physical timescales. The migration timescale of the inner ``test particle" outwards is approximately $10^6$ times the orbital period of the test particle and the eccentricity damping timescale is chosen to be 100x smaller. Both bodies begin with circular orbits outside the 2:1 resonance. In this case, the criterion for whether or not the instability can set in at the 2:1 resonance is given by $m_2/M_\star\lesssim 5\times10^{-4}$, which is easily satisfied (regardless of whether we use the compact approximation or the exact form).  Indeed, we see this single trajectory undergoing capture and then subsequent escape via overstable librations for many resonances. The dotted colored lines show the predictions for the equilibrium eccentricity of the test particle for each resonance (without making the approximation that $R\approx 1$ or $m\approx m+1$).

In the right set of panels, we show the results of integrations of a series of systems where the outer planet is treated as a test particle. In this case, we reversed everything exactly compared with the left set of panels. The outer ``test particle" has a mass of $10^{-8}$ the mass of the star, while the inner planet is $10^{-5}$ the mass of the star. The migration and damping rate for the inner planet is very long compared with all other physical timescales. The migration timescale of the outer test particle is again approximately $10^6$ times the orbital period of the test particle and the eccentricity damping timescale is chosen to be 100x smaller. 

In the upper right plot, the red curve shows the result when both bodies begin with circular orbits outside the 2:1 resonance, the other colors show the evolution when began just outside of other first order mean motion resonances as labeled. For each, the system is captured into resonance and does not escape. That is, the libration amplitude does not grow with time, suggesting that these are in fact stable configurations. If we took the results for the inner test particle case and applied them here, we would expect these systems to easily be unstable as well since $m_1/M_\star < 5\times10^{-4}$. Moreover, if the outer test particle case was analogous to the inner test particle case, the timescale for escape would be much shorter than our integration time, as it was in the case when the inner planet was a test particle. This increases our confidence that we have integrated these systems long enough to show that they are stable (i.e. that the instability time is not significantly longer than the integration time, giving the illusion of stability), and that there truly is a difference between whether the inner planet is a test particle or the outer planet is a test particle.

	\begin{figure}
	\begin{center}
	\includegraphics[width=2.5in]{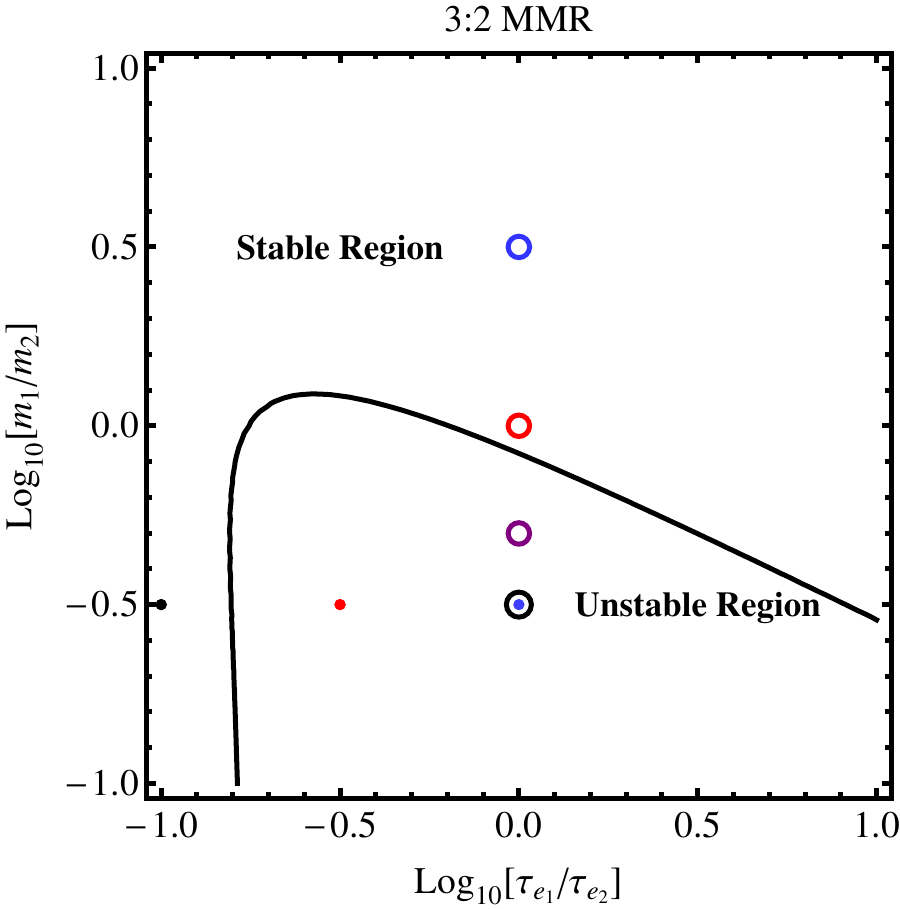}
	\caption{The critical curve (black) showing where overstable librations of the 3:2 resonance can occur for a system of two planets of total mass $\epsilon_p = 1.31\times 10^{-5}$. The evolution of the three systems shown with constant $\zeta$ (filled circles) is shown in Figure \ref{fig:tau_e_mA}, while the evolution of the four systems shown with constant $\tau_{e,1}/\tau_{e,2}$ (open circles) is shown in Figure \ref{fig:tau_e_mB}. The color of the points corresponds to the color of the trajectories in Figure \ref{fig:tau_e_mA} and Figure \ref{fig:tau_e_mB}.  }
	\label{fig:3to2}
	\end{center}
	\end{figure}

	\begin{figure}
	\begin{center}
	\includegraphics[width=3.5in]{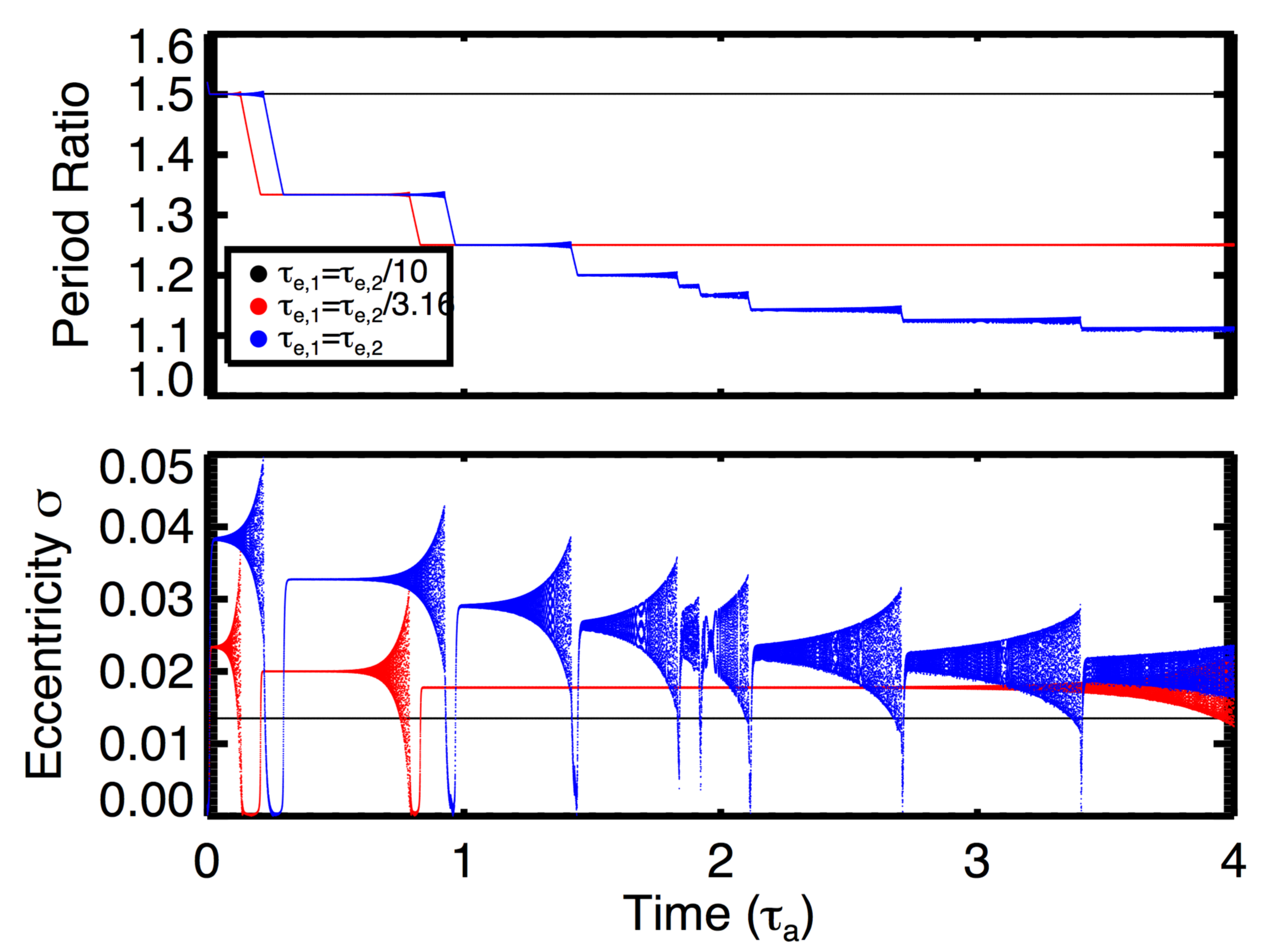}
	\caption{The evolution of three systems with $\epsilon_p = 1.31\times 10^{-5}$, $\zeta = 10^{-0.5}$, $\tau_{a,2}/P_1 \sim 10^6$, $\tau_{a,1} = 5 \tau_{a,2}$, and $\tau_{e,2} = \tau_{a,2}/100$. These trajectories, and their color, correspond to the three filled dots shown at constant $\zeta$ in Figure \ref{fig:3to2}. At this mass ratio, as predicted, only systems with $\tau_{e,1}\gtrsim10^{-0.8} \tau_{e,2} = 0.15\tau_{e,2}$ undergo overstable librations.}
	\label{fig:tau_e_mA}
	\end{center}
	\end{figure}
	
			\begin{figure}
	\begin{center}
	\includegraphics[width=3.5in]{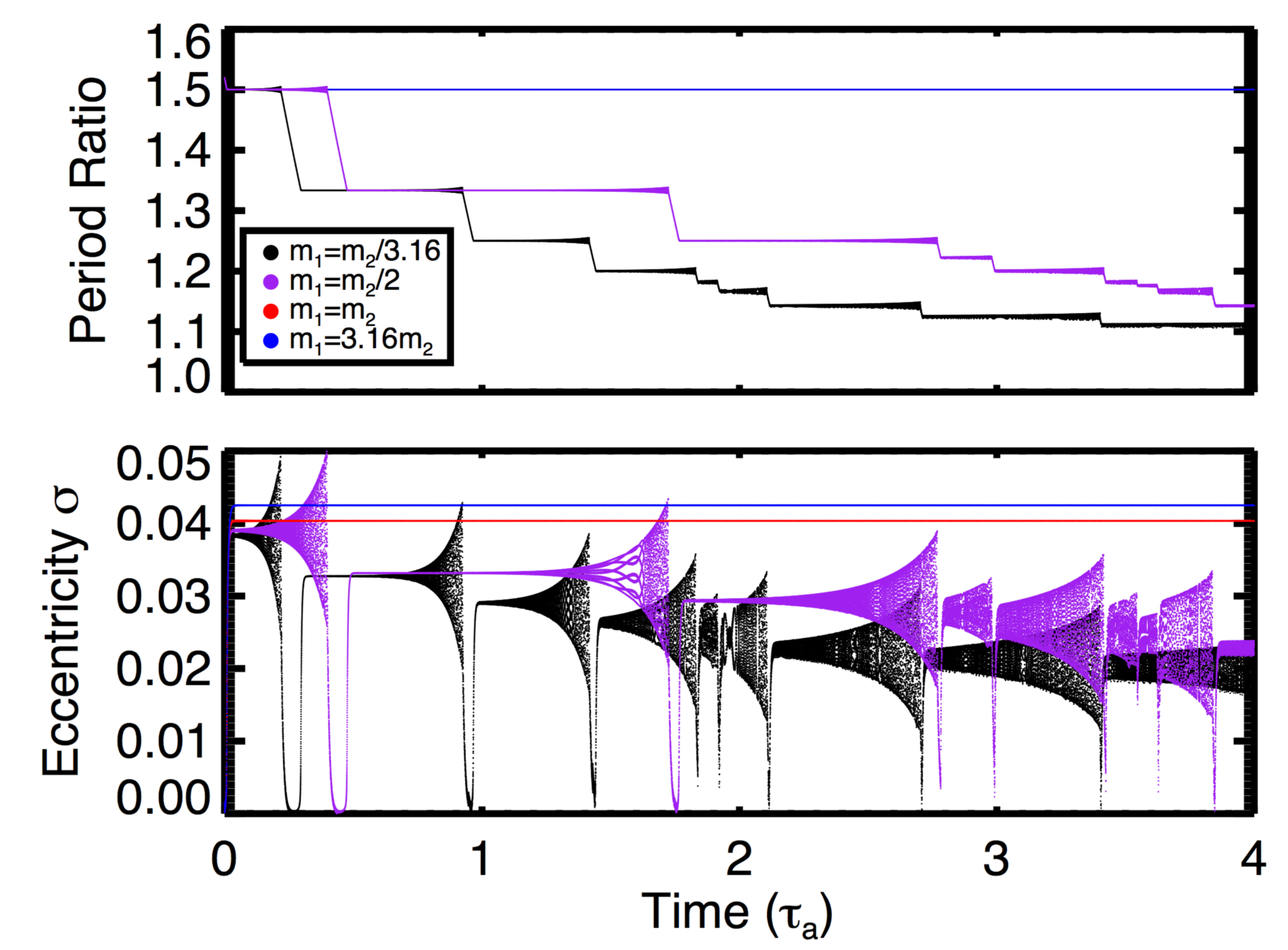}
	\caption{The evolution of four systems with $\epsilon_p = 1.31\times 10^{-5}$, $\tau_{a,2}/P_1 \sim 10^6$, $\tau_{a,1} = 5 \tau_{a,2}$, $\tau_{e,2} = \tau_{a,2}/100$, and $\tau_{e,1} = \tau_{e,2}$. These trajectories, and their color, correspond to the four open circles shown at constant $\tau_{e,1}/\tau_{e,2}$ in Figure \ref{fig:3to2}. At this ratio, as predicted, only systems with $m_{1}\lesssim m_{2} $ undergo overstable librations. (note that the red curve lies below the blue curve in terms of period ratio evolution in the upper panel; the two are distinct in the lower plot showing equilibrium eccentricity, however).}
	\label{fig:tau_e_mB}
	\end{center}
	\end{figure}

We now turn to some tests of the criterion for comparable mass planets. We set $\tau_{a,2}/P_1 \sim 10^6$, $\tau_{a,1} = 5 \tau_{a,2}$, and $\tau_{e,2} = \tau_{a,2}/100$. We then varied $\tau_{e,1}$ and the ratio of $\zeta = m_1/m_2$.  In Figure \ref{fig:3to2}, we show on a panel of $\tau_{e,1}/\tau_{e,2}$ and $m_1/m_2$ where overstable librations can occur for these parameters. The points on this plot correspond to systems we studied migrating convergently beginning from just outside the 3:2 resonance.  We show the evolution of the three systems with a fixed mass ratio $\zeta=10^{-0.5}$ (marked with filled circles in Figure \ref{fig:3to2}) in Figure \ref{fig:tau_e_mA}. As predicted, the two systems with larger values of $\tau_{e,1}/\tau_{e,2}$ (those in red and blue) undergo instability and escape from resonance on a timescale short compared to the integration time, while the system with the smallest value of $\tau_{e,1}/\tau_{e,2}$ (in black) remains in resonance. Note that all of these systems have a value of $\tau_{a,e}>0$. 

In Figure \ref{fig:tau_e_mB}, we show the evolution of four systems marked in Figure \ref{fig:3to2} with open circles with $\tau_{e,1} = \tau_{e,2}$ but varying mass ratio. In this case, systems with $\zeta \gtrsim 1$ should be stable against overstable librations because the quantity $\tau_{a,e}$ is negative. This is indeed what is observed. If $\zeta\lesssim 1$, the orbits may be unstable, but only if the total mass of the planets is low enough.  For the orbits shown in purple and black, that requirement is satisfied, as shown in Figure \ref{fig:3to2} (the open circles in purple and black lie below the solid black curve), and these orbits do escape from resonance.

\section{Discussion}\label{sec:discussion}
 \subsection{Timescales for escape}\label{sec:timescales}
The timescale on which the pair of planets escapes from resonance after reaching an equilibrium eccentricity larger than the critical value is an important quantity because when compared with the migration time $\tau_a$ it will determine what fraction of the time any given pair is found in a resonance vs. migrating between them. The evolution shown in Figure \ref{fig:testPart}, Figure \ref{fig:tau_e_mA} and Figure \ref{fig:tau_e_mB} indicate that the systems which do escape from resonance typically spend the majority of their time trapped in resonances undergoing overstable librations rather than in between them. 

The real (positive) part of the eigenvalue  $\alpha_1$ gives the rate at which orbits spiral away from the fixed point when the orbits are near the fixed point (as it is a local stability analysis). Since the real part of the eigenvalue $\alpha_1$ passes from negative to positive on the critical curve $\epsilon_c=\epsilon_{p,crit}$, the instability time, given by the inverse of the eigenvalue, is very long near the critical curve. The instability time is not the same order of magnitude as $\tau_e$ because the eigenvalue is proportional to $1/\tau_e$ {\it and} the difference of comparable quantities (unless the system satisfies $\epsilon_p \ll \epsilon_{p,crit}$) or vice versa). On the other hand,  the timescale associated with the negative eigenvalue $\alpha_0$, which is also proportional to $1/\tau_e$, is much shorter than timescale associated with the positive eigenvalue $\alpha_1$ since it does not have this dependence on a difference between like quantities. In fact, temporary capture necessitates that the instability timescale be much longer than $\sim 1/|\alpha_0|$.  If the two timescales were comparable, the system would not even be captured into resonance in the first place. This implies that escape from resonance must occur on timescales significantly longer than $\tau_e$ in general.

\begin{figure}
	\begin{center}
	\includegraphics[width=3.5in]{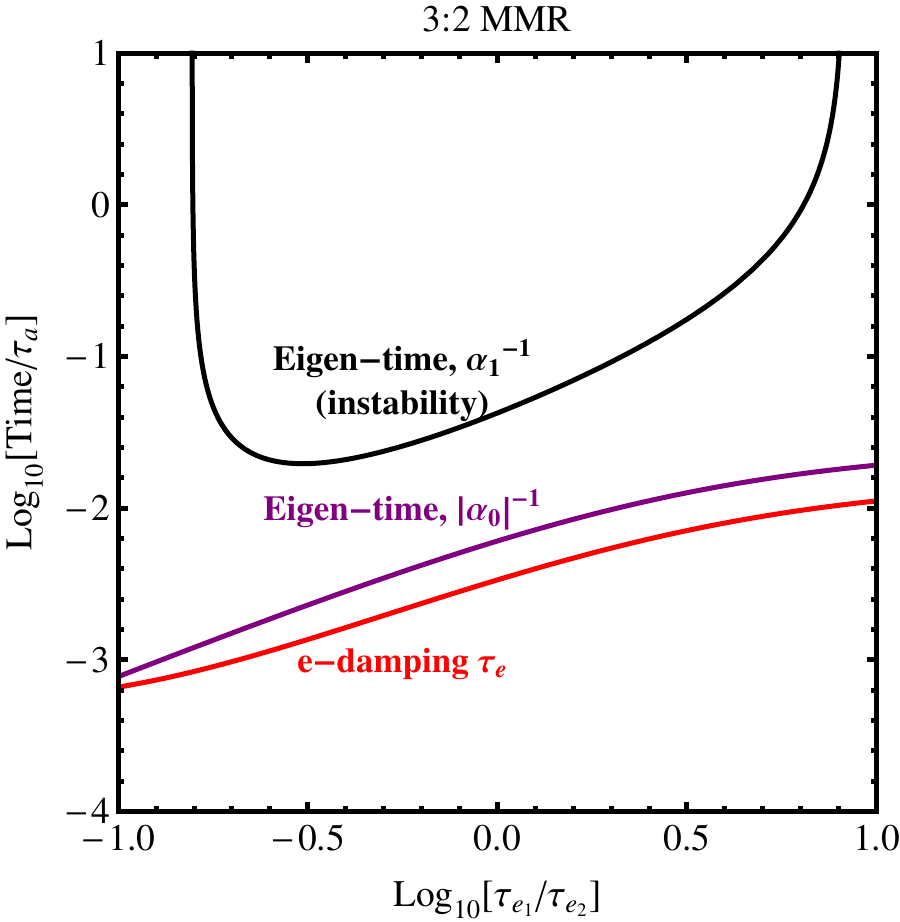}
	\caption{Relevant timescales to the problem of dissipative evolution near the 3:2 mean motion resonance. Here we show $\tau_e$ and the two timescales associated with the real eigenvalues determined by a  linear stability analysis near the fixed point in units of $\tau_a$. The eigen-timescale $1/\alpha_1$ is only shown when $\alpha_1>0$ and the fixed point is ultimately unstable. The relevant parameters used are given in the text.}
	\label{fig:timescales}
	\end{center}
	\end{figure}

To illustrate these points, we show in Figure \ref{fig:timescales} the timescales $\tau_e$, $\alpha_1^{-1}$, and $|\alpha_0|^{-1}$, in units of $\tau_a$,  for the 3:2 resonance for a range of $\tau_{e,1}/\tau_{e,2}$. The inner orbital period is $\approx 10^2$ days. All parameter values correspond to  those of Figure \ref{fig:3to2} except $\zeta$ is fixed at $10^{-0.5}$. One can see that $\alpha_1^{-1}$ (shown only when positive) diverges when the critical curve is reached at Log[$\tau_{e,1}/\tau_{e,2}]\approx -0.8$ and at Log[$\tau_{e,1}/\tau_{e,2}]\approx 0.9$ as in Figure \ref{fig:3to2}. Moreover, across the entire range,  $\tau_e$ is comparable to $|\alpha_0|^{-1}$ and at least an order of magnitude smaller than $\alpha_1^{-1}$. The numerical simulations show that the pairs spend most of their time in resonance, which suggests that approximately 5-10 instability times are required for escape (the former implies roughly $\tau_{escape} \sim \tau_a$ and Figure \ref{fig:timescales} indicates that $\tau_a \sim 10\alpha_1^{-1}$, and therefore $\tau_{escape}\sim 10\alpha_1^{-1}$

  \subsection{Planet pairs discovered with Kepler}
  According to \citet{Sch}, overstable librations of first order resonances could account for the observed period ratio distribution because 1) the typical masses of the Kepler planets are small enough, given the estimated ratio of migration and eccentricity damping rates expected from Type I migration, that the first order resonances are unstable and any capture was only temporary, and 2) the time spent in resonance was small compared to the time spent in between resonances, such that when planet-disk interactions stopped, most pairs were left in between resonances. At the same time, this theory is consistent with giant planets being found in resonance \citep{WrightRV}, since in that case the total mass of the planets is too large compared to the critical value for instability.

In the more general case of two massive planets, we have shown that overstable librations can only occur in specific cases. First, the resonance is stable if the quantity $\tau_{a,e}$ is negative, regardless of the total mass of the planets or the ratio of eccentricity damping to semimajor axis damping, that is, the resonance is stable if
\begin{align}\label{stability_crit}
\frac{\tau_{e,2}}{\tau_{e,1}}< \frac{m_1^2}{m_2^2}\alpha_{\rm{res}}/R^2
\end{align}
 where $R$ and $\alpha_{\rm{res}}$ are of order unity.  If we employ the ``compact approximation" and the scaling that $\tau_{p} \propto \frac{1}{m_p}$, where $p$ stands for planet and $\tau$ a damping timescale due to interaction with the disk (e.g. \citet{GoldreichTremaine,Tanaka1}), we find that resonances are nominally stable to overstable librations if
\begin{align}\label{eq:stability}
m_2\lesssim m_1
 \end{align} 
 Note that for the 2:1 resonance there is considerably more leeway in this mass criterion, as now the pre-factor on the right-hand side of Equation \eqref{stability_crit} evaluates to $\approx 1/12$. For comparison, at the 3:2 resonance it is $1/0.85$. This implies that if  $\tau_{p} \propto \frac{1}{m_p}$,  the 2:1 resonance is stable for 
 \begin{align}
 m_2 \lesssim  m_1/12,
 \end{align}
 regardless of the total mass of the planets.

If the system does fail Equation \eqref{stability_crit}, we find a similar result to that of \citet{Sch}. That is, the system can escape from resonance via overstable librations only if the total mass of the planets is small compared to a critical value. The critical value depends on the ratio of $\tau_a/\tau_e$, $\zeta$ and the relative eccentricity damping times. Therefore, a pair containing a giant planet (regardless of planet mass ratio) will be more likely to be caught in resonance permanently given the high value of $\epsilon_p$, while low mass planet pairs like those found by {\it Kepler} will be liable to escape from resonance.  What this means is that escape from resonance should have occurred for a subset of the {\it Kepler} systems. In particular, the lack of pairs caught in the 2:1 mean motion resonance can be explained through overstable librations assuming the total mass of the planets is small enough and that $m_2\lesssim m_1/12$.  However, unless the inner planet is less massive than the outer planet, capture into all other first order resonances is predicted to be permanent.

 More importantly, even for pairs which undergo overstable librations, the cumulative time that the pair spends in any resonance is much longer than that spent in  between resonances. Therefore the average period ratio distribution of a population of pairs undergoing convergent migration will be peaked near resonances, since all pairs will either 1) be too massive to undergo instability or not satisfy the mass ratio $\zeta$ requirement ($m_1\lesssim m_2$), and therefore be captured permanently into resonance or 2) satisfy the total mass and mass ratio requirement for instability but spend most of their time in a resonance regardless. Overstable librations alone cannot explain the period ratio distribution.

One possible resolution to this issue could be orbital instability. If any pair in a system satisfies the requirements for overstability, that pair will undergo successive repetitions of the following sequence: capture into resonance, followed by overstable librations and escape from resonance. This would lead a gradual compactification of the orbits - and could explain the presence of some very close pairs of planets, like Kepler-36 \citep{Josh}. At some point, if this process continues, the pair will be close enough that when they escape the current resonance, at the equilibrium eccentricity of the resonance, they are in the chaotic region corresponding to overlap of resonances (e.g. \citet{Wisdom1980,ResOverlap}). If the timescale to develop crossing orbits in this chaotic region is short enough, these planets will scatter or collide with each other. If they do not collide, and merely scatter, they will begin to undergo the entire process again, as long as the disk is present. 

When the disk finally dissipates, the planetary system will be left in a precarious position, with compact pairs of planets in resonance with nonzero eccentricities of several percent (corresponding to the equilibria at the center of resonance). Pairs that were undergoing overstable librations will have nonzero libration amplitudes. We hypothesize that these configurations will typically have lifetimes short compared with the ages of the systems, and that many of them will undergo orbital instability, leading to wider distribution of period ratios and pairs not near resonance, in correspondence with the period ratio distribution. This would also explain why there is a decrease in the number of very compact pairs as well, since those orbits would be most unstable (see also \citealt{PuWu}).

\subsection{Effects of external sources of precession}
    The analysis we have undertaken applies when the motion of the planets is governed by the Hamiltonian given in Equation \eqref{eqn:fullHam} and the dissipation prescription of Equations \eqref{damping}. This formulation neglects apsidal regression of the pericenters induced by the gravitational potential of the disk \citep{Heppen,Tamayo} and precession of the pericenters directly caused by the density waves excited by the planet in the disk \citep{Tanaka2}. If these ``forced" precession rates differ between the two planets, the conservative Hamiltonian is no longer integrable \citep{Maryame}. Instead, both resonant angles $\theta_1=\theta-\varpi_1$ and $\theta_2  =\theta-\varpi_2$ are independent. The resonant centers, where $\theta_1$ and $\theta_2$ have zero time derivative, are separated by an amount of $\dot{\varpi}_2-\dot{\varpi}_1$. This splitting changes the period ratio which corresponds to the resonance as $P_2/P_1 = (m+1)/m-\dot{\varpi_i}P_2/(2\pi m)$. 
    
Although a study of the effect of these external precession rates is beyond the scope of this paper, we can provide some general hypotheses as to how they might affect our results. First, in the limit of $\alpha=a_1/a_2 \rightarrow 1$, the precession rates caused by the gravitational potential of the disk - which are independent of the planet masses and but depend on the semimajor axis of the planets - will be approximately equal. If these dominate the external precession rates, there will be no splitting between the resonant angles $\theta_1$ and $\theta_2$, and the integrable analysis should still apply. 

Second, if the precession rates cause a large enough splitting between the two resonances and they do not perturb each other too strongly, they can be treated individually. In this case, the motion would still be integrable near each resonance. If, for example, the system was near the resonance associated with $\theta_1$, we could average the Hamiltonian over the angle $\theta_2$, and we would be left with something akin to the circular restricted three body problem for an inner test particle. In the opposite case, we would be left with a problem akin to the circular restricted three body problem for an outer test particle. As demonstrated by \citet{Sch}, the CR3BP with an inner test particle is susceptible to overstable oscillations, and as shown here, the opposite is not.

As the system evolves, the period ratio will approach the value $(m+1)/m$ from a larger value. The system will first encounter one of the resonances (which it encounters first depends on the precession rates). If the migration rate is slow enough and the eccentricities low enough, the system will be captured. If this first resonance is the $\theta_2$ one, the system will be trapped permanently. However, if the system first encounters the $\theta_1$ resonance, the capture may be temporary if the system satisfies the criterion on $\epsilon_p$. The system could then escape the resonance - but it would then encounter the $\theta_2$ resonance. Any capture again would likely be permanent.

If the splitting between the two resonances is not large enough to treat them individually, the motion may be chaotic, due to overlap of resonances \citep{Maryame}. In this instance the probability of capture and of escape via overstable librations might be best assessed through numerical experiments. 
    
    \section{Conclusion}
    In order to understand the observed sample of exoplanets, we need a better understanding of how various physical processes shape planetary systems after formation. We have built upon the work of \citet{Sch} to account for the effects of eccentricity dependent semimajor axis evolution for a system of two massive planets with mildly eccentric orbits undergoing convergent migration near a first order mean motion resonance.

    We have shown that in the full problem resonances are in general more stable against overstable oscillations compared with the restricted case considered by \citet{Sch}. In particular, for first order resonances other than the 2:1 resonance, if $\tau_{e,2}/\tau_{e,1}\lesssim m_1^2/m_2^2 $, resonance capture is permanent (sans other disruptive effects like turbulence), regardless of the total mass of the planets. If the timescale for damping is inversely proportional to the mass of the planet, then this criterion states that the resonances are stable if $m_2 \lesssim m_1$. For the 2:1 resonance, however, the same is true only if the inner planet is significantly ($\sim 12$x) more massive than the outer. 
    
    Now, if the relative eccentricity damping rates and the mass ratio between the planets do satisfy the criterion $\tau_{e,2}/\tau_{e,1}\gtrsim m_1^2/m_2^2 $, we find a result very similar to that found in the restricted case: resonances are unstable if the total mass of the planets is smaller than a critical value dependent on the generalized $\tau_a$ and $\tau_e$ defined above. This is consistent with the fact that pairs of gas giants have been found in resonance (e.g. \citealt{WrightRV}).

We have also demonstrated that in the case where the resonances are unstable the timescale to escape from resonance is not simply proportional to $\tau_e\ll \tau_a$. Instead, the instability timescale is much longer than $\tau_e$, and furthermore $\sim$5-10 instability times are required for escape. Because of this, a pair of planets undergoing overstable oscillations spends the majority of time in a resonance rather than in migrating between them.

It is important to determine if the lack of period commensurabilities within the  {\it Kepler} data is evidence that convergent migration did not occur. This paper has attempted to address whether the mechanism proposed by \citet{Sch} can explain the lack of near resonant pairs within the context of orbital migration. Testing predictions about which {\it Kepler} pairs could have escaped from resonance based on their mass ratio is difficult since the {\it Kepler} planets do not typically have measured masses, and the radius ratio may not be a reliable proxy \citep{Weiss}. However, based on the timescale argument, even if the majority of pairs  did satisfy the criterion for overstability,  the lack of near resonant pairs within the {\it Kepler} period ratio distribution is not due to pairs simply spending more time between resonances than trapped in them, so that when the gas disk dissipated more pairs were left between resonances than near them.  From this we conclude that overstable librations, which should have occurred for some pairs, cannot single-handedly account for the lack of pairs near resonance.

We have argued that regardless of whether planets form in-situ or at larger orbital distances, they likely interacted with a gaseous disk, the presence of which may be required to explain the volatile rich nature of the exoplanets larger than $1.6R_\oplus$. In this case, then, a loss of orbital energy and some exchange of angular momentum with the disk is unavoidable, and therefore migration must be brought into agreement with the lack of near resonant pairs.  An interesting idea left for further investigation is that the effects of higher multiplicity systems and of true orbital instability could account for the period ratio distribution in combination with overstable librations.
\acknowledgments
K.M.D. would like to acknowledge support from the Joint Center for Planetary Astronomy at Caltech and also to thank Peter Goldreich for prompting the discussion of external sources of precession.

\bibliographystyle{apj}
\bibliography{overstable7}

\clearpage
 \renewcommand{\theequation}{A.\arabic{equation}}    
  \setcounter{equation}{0}  
    \renewcommand{\thesubsection}{A-\arabic{subsection}}   
     \setcounter{subsection}{0}  
        \renewcommand{\thefigure}{A.\arabic{figure}}   
          \setcounter{figure}{0}  
  \section*{APPENDIX}\label{sec:HamDevelop}  
\subsection{Setup of integrable problem}
We consider a system of two planets of mass $m_1$ and $m_2$ orbiting a star of mass $M_\star$ with periods near the $m$:$m+1$ first order mean motion resonance. We assume the orbits are nearly circular and nearly coplanar, and at this stage we ignore external sources of pericenter precession which arise from the gravitational potential of the protoplanetary disk as well as from direct interactions with the disk.  In this configuration, the Hamiltonian can be reduced to a one-degree of freedom system with a single free parameter \citep{Sessin}. This is because, after expanding the Hamiltonian about the resonance center, there is a series of canonical transformations which reduce the number of degrees of freedom from 4 to 1 \citep{WisdomCtB,Lemaitre}.

Before presenting that simplified Hamiltonian, we motivate it as follows. As given in Equation \eqref{eqn:fullHam}, the Hamiltonian near the $m$:$m+1$ resonance takes the following form:
\begin{align}\label{eqn:fullHamAgain}
H & = -\frac{GM_\star m_1}{2 a_1}-\frac{GM_\star m_2}{2 a_2}-\frac{Gm_1m_2}{a_2}\times \nonumber \\
&\bigg[ f_{m+1,27}(\alpha_{\rm{res}})e_1 \cos{[\theta-\varpi_1]} +[f_{m+1, 31}(\alpha_{\rm{res}})-\delta_{m,1}2\alpha_{\rm{res}}]e_2 \cos{[\theta-\varpi_2]}  \bigg]
\end{align}
where $\theta = (m+1)\lambda_2 -m\lambda_1$ and $a_i$, $e_i$, $\lambda_i$ and $\varpi_i$ are the semimajor axes, eccentricities, mean longitudes, and longitudes of periastron of the two planets. The quantities $f_{j,27}$ and $f_{j,31}$ are \citep[p.~539-556]{MurrayDermott}
 \begin{align}
 f_{j,27}(\alpha) &= \frac{1}{2}(-2 j-\alpha \frac{d}{d\alpha}) A_j(\alpha) \nonumber \\
 f_{j,31}(\alpha) &= \frac{1}{2}(-1+2j+\alpha \frac{d}{d\alpha}) A_{j-1}(\alpha)\nonumber \\
 A_j(\alpha) & = \frac{1}{\pi} \int_0^{2\pi} \frac{\cos{(j \phi)}}{\sqrt{1-2\alpha \cos{\phi}+\alpha^2}}d \phi.
 \end{align} 
After expansion about the resonance center, these coefficients are evaluated at $\alpha \equiv a_1/a_2 =\alpha_{\rm{res}}$, where $\alpha_{res}  = [m/(m+1)]^{2/3}$. Treated as functions of $m$, these coefficients are well approximated as 
\begin{align}\label{laplace_fit}
f_{m+1,27} & \approx -0.8 m \nonumber \\
f_{m+1,31} & \approx 0.8 m
\end{align}
as shown by \citet{Quillen2011} (and easily confirmed numerically).

The osculating elements are not canonical. Instead we  use the following as our variables:
 \begin{align}\label{var}
\Lambda_i &= m_i \sqrt{G m_\star a_i}  \nonumber \\
\lambda_i  &= \Omega_i+\omega_i+M_i \nonumber \\
P_i &= m_i \sqrt{G m_\star a_i}  (1-\sqrt{1-e_i^2})  \approx \Lambda_i \frac{e_i^2}{2}\nonumber \\
p_i &= -\varpi_i = -\Omega_i-\omega_i
 \end{align} 
At first order in the inclinations $I_i$, the Hamiltonian is independent of $\Omega$, the longitude of ascending node, and $I$. This implies that the inclinations need to be large enough for $I^2$ terms to be important for non-coplanarity to have an effect on the orbits. Therefore, the nearly coplanar regime is also well described by our formulation even though inclinations do not appear in the Hamiltonian.
 
 Instead of the polar set $(P,p)$, we use the cartesian one defined as
 \begin{align}
  x_i & = \sqrt{2 P_i} \cos{p_i} \nonumber \\
  y_i & = \sqrt{2 P_i} \sin{p_i}
 \end{align}
 which is more appropriate in the limit of low eccentricities. Then the perturbation Hamiltonian $H_1$ can be written as 
 \begin{align}
 H_1 \propto \frac{1}{\Lambda_2^2}\times &\bigg[ \bigg(\frac{f_{m+1,27}(\alpha_{\rm{res}})}{\sqrt{\Lambda_1}} x_1 +\frac{f'_{m+1, 31}(\alpha_{\rm{res}})}{\sqrt{\Lambda_2} } x_2\bigg) \cos{\theta} -  \bigg(\frac{f_{m+1,27}(\alpha_{\rm{res}})}{\sqrt{\Lambda_1}} y_1 +\frac{f'_{m+1, 31}(\alpha_{\rm{res}})}{\sqrt{\Lambda_2} } y_2\bigg) \sin{\theta} \bigg]
  \end{align}
 where for simplicity $f'_{m+1, 31}(\alpha_{\rm{res}}) = [f_{m+1, 31}(\alpha_{\rm{res}})-\delta_{m,1}2\alpha_{\rm{res}}]$. Moreover, after expanding about the resonance center, all $\Lambda_i$ are held fixed at their resonant values $\Lambda_{i,\rm{res}}$. As messy as this looks, then, the coefficients of the $\cos{\theta}$ and $\sin{\theta}$ terms are simply linear combinations of $x_i$ and $y_i$. With an appropriate normalization, these combinations are a canonical rotation of the original cartesian eccentricity variables. In fact, the coefficient of the $\cos{\theta}$ term is a canonical momentum to a coordinate equal to the coefficient of the $\sin{\theta}$ term. This is the key - this rotation takes the 2 degrees of freedom associated with the eccentricity of each planets and reduces it to one degree of freedom (the conserved quantity generated here is $\Psi_2$ and defined below). 
 
 Furthermore, since each $\lambda_i$ only appears as the combination $\theta$, there is a further reduction in the number of degrees of freedom and an associated conserved quantity which we refer to as $\Theta_1$. The action associated with $\theta$ is denoted as $\Theta$, and,
   \begin{align}
   \Theta_1 &= \frac{m}{m+1}\Lambda_2+\Lambda_1 \nonumber \\
  \Theta &= \Lambda_2/(m+1)
  \end{align}
  (see e.g. \citealt{ResOverlap} for details of these canonical transformations). We then choose units, such that actions are measured in terms of  $\Theta_1$, the Hamiltonian in terms of  $\Theta_1/n_{\Theta_1}$, where 
\begin{align}
n_{\Theta_1} &= \frac{n_2}{\bigg(\frac{m}{(m+1)}+\zeta\sqrt{\alpha_{\rm{res}}}\bigg)^3}
\end{align}  
and such that time is measured in units of $1/n_{\Theta_1}$. That implies that derivatives of the Hamiltonian, which result in equations of motion for the variables, are time derivatives with respect to the unitless time $\hat{t} =tn_{\Theta_1}$. Hats will denote this set of variables.

We define
 \begin{align}
  \hat{x}_i & = x_i/\sqrt{\Theta_1} \nonumber \\
 \hat{y}_i & = y_i/\sqrt{\Theta_1} \nonumber \\ 
    \bar{\delta}_1 & =\frac{1}{\Lambda_{2,\rm{res}}^2} \frac{f_{m+1,27}}{\sqrt{\Lambda_{1,\rm{res}}}} \nonumber \\
\bar{ \delta}_2 & = \frac{1}{\Lambda_{2,\rm{res}}^2} \frac{f_{m+1,31}}{\sqrt{\Lambda_{2,\rm{res}}}} \nonumber \\
\bar{\delta} & = \sqrt{\bar{\delta}_1^2+\bar{\delta}_2^2} \nonumber \\
 r_1 & = \frac{\bar{\delta}_1 \hat{x}_1 +\bar{\delta}_2 \hat{x}_2}{\bar{\delta}} \nonumber \\
 s_1 & = \frac{\bar{\delta}_1 \hat{y}_1 +\bar{\delta}_2 \hat{y}_2 }{\bar{\delta}} \nonumber \\
  \Phi & = \frac{1}{2}(r_1^2+s_1^2) \nonumber \\
\tan{\psi} & = \frac{s_1}{r_1}
\end{align}
 where coefficients of the rotation alluded to above can be read off the Hamiltonian and are given by $\bar{\delta}_1$ and $\bar{\delta}_2$. The rotated combinations are $r_1$ (``momentum") and $s_1$ (``position"). 
 The momentum $\Phi$ and angle $\psi$ reflect a polar canonical transformation of the cartesian variables $r_1$ and $s_1$. $\Phi$ is related to the eccentricity as
 \begin{align}\label{eqn:phisigmarelation}
 \Phi & = \frac{m+1}{m}\frac{\alpha_{\rm{res}}}{\alpha_{\rm{res}}+\zeta}\frac{\zeta\sqrt{\alpha_{\rm{res}}}}{2(R^2+\zeta\sqrt{\alpha_{\rm{res}}})}\sigma^2 \nonumber \\
 \sigma^2 & = (R^2e_1^2+e_2^2-2Re_1e_2\cos{\Delta \varpi}) \nonumber \\
 \zeta & = \frac{m_1}{m_2} \nonumber \\
 R & = \frac{\abs{f_{m+1, 27}(\alpha_{\rm{res}}) }}{f'_{m+1, 31}(\alpha_{\rm{res}}) }
 \end{align}
 
 After the rotation and change to the polar variables $\Phi$ and $\psi$ has been performed,  only a single combination of the remaining angles $\psi$ and $\theta$ appears in the Hamiltonian. A final canonical transformation to this angle and a its canonical action (equal to $\Phi$) is performed. The last canonical transformation yields a single degree of freedom Hamiltonian\footnote{In the \citet{ResOverlap} paper, the following Hamiltonian is defined as $K$ and given in Equation 26.} of the form
 \begin{align}\label{Ham6}
\hat{H} & = \frac{1}{2}{\beta} (\Phi-\Gamma)^2 - \epsilon_1 \bar{\delta} \sqrt{2 \Phi} \cos{\phi},
\end{align} 
where 
\begin{align}
\phi &= \theta+\psi \nonumber \\
\beta &= -3 \frac{m^2 (\alpha_{res}+\zeta)^5}{ \zeta \alpha_{res}} \nonumber \\
\epsilon_1 & = \frac{m_1}{M_\star} \nonumber \\
 \Gamma & =  \Phi-\delta \Theta \nonumber \\
 \delta \Theta &= \hat{\Theta}-\bar{\Theta} \nonumber \\
 \bar{\Theta} & = \frac{\alpha_{res}}{m(\alpha_{res}+\zeta)}
\end{align}

The quantity $\Gamma$ is conserved (it is the conserved quantity found after realizing that only the combination $\theta+\psi$ appears, and not each individually; refer to e.g. \citealt{ResOverlap}). When $\delta \Theta =0$, or $\Theta = \bar{\Theta}$, the system is at the exact period commensurability and $\alpha = \alpha_{res}$. Finally, we turn now to the conserved quantity denoted $\Psi_2$, found after performing the rotation. It can be written in terms of the Poincare variables as
\begin{align}
\Psi_2 & = \frac{1}{2}(r_2^2+s_2^2) \nonumber \\
 r_2 & = \frac{\bar{\delta}_2 \hat{x}_1 -\bar{\delta}_1 \hat{x}_2}{\bar{\delta}} \nonumber \\
 s_2 & = \frac{\bar{\delta}_2 \hat{y}_1 -\bar{\delta}_1 \hat{y}_2 }{\bar{\delta}} \nonumber \\
 \tan{\psi_2} & = \frac{s_2}{r_2} \nonumber \\ 
\end{align}

$\Psi_2$ is conserved because the angle $\psi_2$ does not appear in the original Hamiltonian. Note that if we wish to extend this analysis to second order in the eccentricities of the planets, the angle $\psi_2$ appears in the Hamiltonian and $\Psi_2$ is no longer conserved.  These new ``eccentricity" vector components $(r_i, s_i)$ are obtained from a linear transformation of the original $(x_i,y_i)$, and $\Phi +\Psi_2 = (P_1+P_2)/\Theta_1$ since the transformation is a rotation that preserves length. 

At this stage (Hamiltonian \eqref{Ham6}), our variables are the momentum (action) $\Phi$ and the conjugate angle $\phi$.  One final rescaling is performed to put the Hamiltonian into the form given in Equation \eqref{Ham1}.  Primes will refer to scaled quantities. We divide actions by the unitless quantity $Q$. We rescale the Hamiltonian and time $t$ using a second parameter $a$, as $H' = \hat{H}/a$ and $t' = \hat{t}a/Q$. Choosing $a = Q^2 |\beta|$, and $Q = (\epsilon_1 \bar{\delta}/|\beta|)^{2/3}$ leaves us with a single free parameter ($\Gamma'$). The parameter Q is 
\begin{align}
Q & = \epsilon_p^{2/3} \zeta \bigg(\frac{f_{m+1,31}^2}{9(m+1)(1+\zeta)^2}\bigg)^{1/3}\frac{\alpha_{res}^{5/6}}{m} \bigg(\frac{R^2+\zeta\sqrt{\alpha_{res}}}{(\alpha_{res}+\zeta)^5}\bigg)^{1/3},
\end{align}
where we have changed from $\epsilon_1$ to the total mass of the planets $\epsilon_p=\epsilon_1+\epsilon_2$.

The final Hamiltonian is 
\begin{align}
H' & = -\frac{1}{2} (\Phi'-\Gamma')^2 - \sqrt{2 \Phi'} \cos{\phi}.
\end{align}
where the ``time" derivatives of the variables $\Phi'$ and $\phi$ are with respect to $t' =  t n_{\Theta_1}Q|\beta|$.

The conservative system is governed by the equations
\begin{align}
\frac{d \Phi'}{d t'}\bigg\lvert_c & = -\frac{\partial H'}{\partial \phi}  = -\sqrt{2 \Phi'} \sin{\phi}\nonumber \\
\frac{d \phi}{d t'}\bigg\lvert_c & = \frac{\partial H'}{\partial \Phi'}  = -(\Phi'-\Gamma')-\frac{1}{\sqrt{2\Phi'}}\cos{\phi}\nonumber \\
\frac{d\Gamma'}{d t'}\bigg\lvert_c & = 0
\end{align}
where $\bigg\lvert_c$ indicates the conservative evolution.

\subsection{Non-conservative forces}
We now add to the conservative system the effects of eccentricity damping and migration. We parametrize these effects as
\begin{align}
\frac{1}{e_i}\frac{d e_i}{d t'} & = - \frac{1}{Q|\beta| n_{\Theta_1}}\frac{1}{\tau_{e_i}} \nonumber \\
\frac{1}{a_i}\frac{d a_i}{d t'} & = \frac{1}{Q|\beta| n_{\Theta_1}}\bigg(-\frac{ 2p  e_i^2}{\tau_{e_i}} -\frac{1}{\tau_{a_i}}\bigg)
\end{align}
We now  convert these time variations in semimajor axis and eccentricity into time variations of $\Phi'$ and $\Gamma'$. First,
\begin{align}
\frac{1}{\Lambda_i}\frac{d \Lambda_i}{d t'} & = \frac{1}{2a_i}\frac{d a_i}{d t'} = -\frac{1}{Q|\beta| n_{\Theta_1}}\bigg(\frac{ p  e_i^2}{\tau_{e_i}} +\frac{1}{2\tau_{a_i}}\bigg)\nonumber \\
\frac{1}{P_i}\frac{d P_i}{d t'}  & = \frac{1}{\Lambda_i}\frac{d \Lambda_i}{d t'}  + 2 \frac{1}{e_i}\frac{d e_i}{d t'}  = -\frac{1}{Q|\beta| n_{\Theta_1}}\bigg(\frac{p  e_i^2+2}{\tau_{e_i}} +\frac{1}{2\tau_{a_i}} \bigg)
\end{align}

At this point, we make the assumption that $\Psi_2=0$ and remains zero during the evolution of the system around and in resonance. This greatly simplifies the equations for $\Phi'$ and $\phi$ as we will show immediately below. That $\Psi_2$ is small when the resonance is encountered is consistent with the fact that the orbits are undergoing eccentricity damping and therefore have nearly circular orbits prior to resonance encounter. As long as the eccentricities remain low, the motion is well described by the Hamiltonian at $O(e)$ used here, which is independent of $\psi_2$, and which therefore conserves $\Psi_2=0$. However, if the eccentricities grow too large as the system evolves in the resonance (since capture into resonance excites eccentricities), the $O(e^2)$ terms in the Hamiltonian will become important, and $\Psi_2$ will no longer be conserved at zero. We will discuss when our assumption of $\Psi_2=0$ $\forall$ $t$ breaks down in Section \ref{sec:psi2zero}.

Now, $\Phi +\Psi_2 = (P_1+P_2)/\Theta_1$, and 1) $\Psi_2\approx0$, and 2) it remains so (i.e. the time derivative is small). Therefore, we can write $\Phi = (P_1+P_2)/\Theta_1$ such that
\begin{align}
\frac{d \Phi'}{d t'}\bigg\lvert_{\rm{dis}} & \approx  \frac{1}{\Theta_1 Q} \bigg(\frac{dP_1 +dP_2}{d t'}-\Phi\frac{d \Theta_1}{d t'}\bigg) \nonumber \\
& = \frac{1}{ Q^2|\beta| n_{\Theta_1}}\bigg[\frac{P_1}{\Theta_1} \bigg(\frac{-p  e_1^2-2}{\tau_{e_1}} -\frac{1}{2\tau_{a_1}} \bigg) + \frac{P_2}{\Theta_1} \bigg(\frac{-p  e_2^2-2}{\tau_{e_2}} -\frac{1}{2\tau_{a_2}} \bigg) \nonumber \\
&-\frac{\Phi}{\Theta_1} \bigg( \frac{m}{m+1}\frac{\Lambda_2}{2}\bigg(\frac{ -2p  e_2^2}{\tau_{e_2}} -\frac{1}{\tau_{a_2}}\bigg)  +\frac{\Lambda_1}{2} \bigg(\frac{ -2p  e_1^2}{\tau_{e_1}} -\frac{1}{\tau_{a_1}}\bigg)\bigg) \bigg]
\end{align}

Moreover, $\Psi_2=0$ implies that $\bar{\delta}_2^2 P_1 = \bar{\delta}_1^2 P_2$, and therefore from $\Phi = (P_1+P_2)/\Theta_1$ and $P_i = \Lambda_i e_i^2/2$ we can write
\begin{align}
\frac{P_1}{\Theta_1} & = \bigg(\frac{\bar{\delta}_1}{\bar{\delta}_2}\bigg)^2\frac{\Phi}{1+\bigg(\frac{\bar{\delta}_1}{\bar{\delta}_2}\bigg)^2}\equiv \gamma_1 Q\Phi' \nonumber \\
\frac{P_2}{\Theta_1} & = \frac{\Phi}{1+\bigg(\frac{\bar{\delta}_1}{\bar{\delta}_2}\bigg)^2}  \equiv \gamma_2 Q\Phi' \nonumber \\
e_1^2& =2 \eta_1 Q \gamma_1 \Phi' \nonumber \\
e_2^2& = 2 Q \eta_2  \gamma_2 \Phi' \nonumber \\
\eta_1 & \equiv \frac{\Theta_1}{\Lambda_1}  =\bigg(\frac{m}{m+1}\frac{1}{\zeta \sqrt{\alpha_{res}}}+1 \bigg) \nonumber \\
\eta_2 & \equiv \frac{\Theta_1}{\Lambda_2}  = \bigg(\frac{m}{m+1}+\zeta \sqrt{\alpha_{res}} \bigg)
\end{align}

so the equation for the slow evolution of $\Phi'$ reduces to:
\begin{align}\label{c_defn}
\frac{d \Phi'}{d t'}\bigg\lvert_{\rm{dis}} &=\frac{\Phi'}{Q|\beta| n_{\Theta_1}}( C_0 + C_1 \Phi') \nonumber \\
C_0 & = -\gamma_1\bigg(\frac{1}{2 \tau_{a_1}}+\frac{2}{\tau_{e_1}}\bigg)+\frac{1}{2\eta_1 \tau_{a_1}} -\gamma_2\bigg(\frac{1}{2 \tau_{a_2}}+\frac{2}{\tau_{e_2}}\bigg)+\frac{m/(m+1)}{2\eta_2 \tau_{a_2}} \nonumber \\
C_1 & = 2 p Q \bigg( \frac{\gamma_1(1-\gamma_1 \eta_1)}{\tau_{e_1}} +\frac{\gamma_2(m/(m+1)-\gamma_2 \eta_2)}{\tau_{e_2}}\bigg)
\end{align}

Provided $\Psi_2 = 0$, the angle $\phi = (m+1)\lambda_2-\lambda_1+\psi$ has no time evolution due to non-conservative forces. The mean longitudes are independent variables that are unaffected by the slow evolution of $a$ and $e$. However, the angle $\psi$ does depend on $a_i$ and $e_i$ explicitly, through its dependence on $\hat{x}_i$ and $\hat{y}_i$. First, $\Psi_2=0$ implies that $p_1-p_2 = \pi$,  $\tan{\psi} = \tan{p_1}$ and $\tan{\psi}=\tan{(p_2+\pi)} = \tan{p_2}$ (recall $\bar{\delta}_1$ is negative while $\bar{\delta}_2$ is positive). From the definition $\psi =\arctan{s_1/r_1}$, we can derive
\begin{align}
d(\tan{\psi}) & = \frac{d(s_1)}{r_1}-\tan{\psi}\frac{d(r_1)}{r_1} \nonumber \\
& = \frac{\bar{\delta}_1 d(y_1) + \bar{\delta}_2 d(y_2)}{r_1} -\tan{\psi}\frac{\bar{\delta}_1 d(x_1) + \bar{\delta}_2 d(x_2)}{r_1} \nonumber \\
& = \frac{\bar{\delta}_1}{r_1}\bigg[1+\bigg(\frac{\bar{\delta}_2}{\bar{\delta}_1} \bigg)^2 \bigg]\bigg( d(y_1)-\tan{\psi}d(x_1)\bigg)
\end{align} 
but $d(y_i)/y_i  = d(x_i)/x_i$, so the quantity in parentheses is zero. Therefore the derivative of $\psi$ is zero as well:
\begin{align}
\bigg( d(y_1)-\tan{\psi}d(x_1)\bigg) & = d(y_1)\bigg(1-\frac{\tan{\psi}}{\tan{p_1}}\bigg)= d(y_1)\bigg(1-1\bigg)=0
\end{align} 

Lastly, we turn to the evolution of the single parameter $\Gamma'$ appearing in the conservative Hamiltonian. 
\begin{align}\label{a_defn}
\frac{d\Gamma'}{dt'}\bigg\lvert_{\rm{dis}} & = \frac{d}{d t'}(\Phi' - \delta \Theta' ) = \frac{d \Phi '}{d t'} -\frac{1}{Q}\frac{d}{d t'} \bigg(\frac{1}{m+1}\frac{\Lambda_2}{\Theta_1}-\bar{\Theta}\bigg) \nonumber \\
& = \frac{d \Phi '}{d t'} -\frac{1}{Q(m+1)}\frac{d}{d t'} \bigg(\frac{\Lambda_2}{\Theta_1}\bigg) \nonumber \\
& = \frac{1}{Q|\beta| n_{\Theta_1}}\bigg(A_0+(A_1+C_0)\Phi'+C_1\Phi'^2\bigg) \nonumber \\
A_0 & = -\frac{1}{2 Q (m+1) \eta_2}\bigg( \frac{1}{\tau_{a_2}}\bigg[ \frac{m}{\eta_2(m+1)}-1\bigg]+\frac{1}{\eta_1 \tau_{a_1}}\bigg) \nonumber \\
A_1 & = -\frac{2p}{m+1}\bigg( \frac{\gamma_2}{\tau_{e_2}}\bigg[ \frac{m}{\eta_2(m+1)}-1\bigg] +\frac{\gamma_1}{\eta_2 \tau_{e_1}}\bigg)
\end{align}

\subsection{Fixed point of the non-conservative system}
In total, the full equations of motion are the sum of the conservative and non-conservative pieces:
\begin{align}\label{EOMFULL}
(\rm{E}1) &\mbox{   }\frac{d \Phi'}{d t'}    = -\sqrt{2 \Phi'} \sin{\phi}+ \frac{\Phi'}{Q|\beta| n_{\Theta_1}}( C_0 + C_1 \Phi') \nonumber \\
(\rm{E}2) &\mbox{   }\frac{d \phi}{d t'}  \mbox{  } = -(\Phi'-\Gamma')-\frac{1}{\sqrt{2\Phi'}}\cos{\phi} \nonumber \\
(\rm{E}3) &\mbox{   }\frac{d\Gamma'}{d t'}  = \frac{1}{Q|\beta| n_{\Theta_1}}\bigg(A_0+(A_1+C_0)\Phi'+C_1\Phi'^2\bigg) 
\end{align}
We note that $Q \propto \epsilon_p^{2/3}$ and therefore $C_1\Phi'^2 \propto e^4/(\epsilon^{2/3}\tau_{e})$, $(C_0+A_1)\Phi' \propto e^2/(\epsilon^{2/3}\tau_{e})$ and $A_0 \propto 1/(K \epsilon^{2/3} \tau_e)$, where $K=\tau_a/\tau_e$. Therefore, we can neglect the $C_1$ term in Equation (E1) above provided that $e^2\ll1$, and as long as  $1/K\gg e^4$, the $C_1$ term in Equation (E3) is also much smaller than the $A_0$ term. From here on, then, we ignore the $C_1$ contribution. 

Setting all of these equations equal to zero allows us to solve for the equilibrium of the system. Equation (E3) implies the fixed point is given by 
\begin{align}
\Phi'_{eq} \approx -\frac{A_0}{A_1+C_0}
\end{align}
Since $\Phi' = \Phi/Q$, $\Phi_{eq} = -Q A_0/(C_0+A_1)$ which is independent of $Q$ since $A_0\propto 1/Q$. This implies that the equilibrium eccentricities are independent of $\epsilon_p$, the total mass of the planets relative to the star. Very roughly, $C_0+A_1 \propto \frac{1}{\tau_e}$, since $\tau_e \ll \tau_a$, while $Q A_0 \propto \frac{1}{\tau_a}$. Therefore, $\Phi \propto e_{eq}^2 \propto \tau_e/\tau_a = 1/K$. This is consistent with dropping the $C_1$ term in Equations (E1) and (E3), which required that $e^2 \ll 1$ and $1/K \ll e^4$. A faster rate of eccentricity damping, relative to the migration, results in a larger $K$, and a smaller equilibrium eccentricity. 

From Equation (E1), we can solve for $\sin{\phi_{eq}}$:
\begin{align}\label{phieq}
\sin{\phi_{eq}} & \approx \frac{\sqrt{2\Phi'_{eq}}}{2 Q|\beta| n_{\Theta_1}} C_0 \propto -\frac{e_{eq}}{\epsilon_p n_{\Theta_1} \tau_e} \propto -\frac{1}{\epsilon_p n_{\Theta_1} \tau_a}
\end{align}
(note $C_0<0$). Lastly, we need the equilibrium value $\Gamma'_{eq}$. It is a good approximation that $\cos{\phi_{eq}}=-1$ because the deviation of $\phi_{eq}$ from $\pi$ is small for slow dissipation ($n_{\Theta_1}\tau_a\gg 1$ in Equation \eqref{phieq}), and the difference between $\cos{\phi_{eq}}$ and -1 is quadratic in this deviation. In that case, Equation (E2) of Equations \eqref{EOMFULL} yields
\begin{align}
\Gamma'_{eq} \approx \Phi'_{eq}-\frac{1}{\sqrt{2\Phi'_{eq}}}
\end{align}

In the conservative problem, $\phi_{eq}=\pi$.  Equation (E2) of Equations \eqref{EOMFULL} does not contain dissipative terms, and therefore the relationship between $\Gamma'_{eq}$ and $\Phi'_{eq}$ is approximately unchanged by the dissipation since $\cos{\phi_{eq}}$ is approximately unchanged as well. That implies that the fixed point of the dissipative problem $(\phi_{eq},\Phi'_{eq},\Gamma'_{eq})$ lies very close to the fixed point of the conservative problem with $(\phi_{eq} = \pi,\Phi'_{eq})$ with a proximity parameter of $\Gamma'= \Gamma'_{eq}$'. This is the fixed point labeled $x_1$ in Figure \ref{fig:contours}.

Lastly, so as not to introduce extra parameters in the main text, we wrote the dissipative terms appearing in the equations for $\Phi'$ and $\Gamma'$ as
\begin{align}
\mbox{   }\frac{d \Phi'}{d t'}\bigg\lvert_{\rm{dis}} &   = \frac{c_0}{\tau_{e}}\Phi' \nonumber \\
 \mbox{   }\frac{d\Gamma'}{d t'}\bigg\lvert_{\rm{dis}} & = \frac{a_0}{\tau_a}+ \bigg[\frac{c_0}{\tau_{e}}+\frac{p a_1}{\tau_{a,e}}\bigg]\Phi' 
\end{align}
where the parameter $a_1$ here has nothing to do with the semimajor axes. By comparison with Equations \eqref{EOMFULL} one finds that
\begin{align}\label{EOM_CONSTANTS}
c_0 & = \frac{C_0 \tau_{e}}{Q|\beta| n_{\Theta_1}} \nonumber \\
a_0 & = \frac{A_0 \tau_{a}}{Q|\beta| n_{\Theta_1}} \nonumber \\
a_1 & = \frac{A_1 \tau_{a,e}/p}{Q|\beta| n_{\Theta_1}} 
\end{align}
 where we had anticipated that the $C_1$ term is negligible.

\subsection{Linear stability analysis of the fixed point}
The matrix whose eigenvalues we wish to determine is

\[M = \left[ \begin{array}{ccc}
\partial(\frac{d\Phi'}{dt'})/\partial \Phi' & \partial(\frac{d\Phi'}{dt'})/\partial \phi  & \partial(\frac{d\Phi'}{dt'})/\partial \Gamma' \\
\partial(\frac{d\phi}{dt'})/\partial \Phi' & \partial(\frac{d\phi}{dt'})/\partial \phi  & \partial(\frac{d\phi}{dt'})/\partial \Gamma' \\
\partial(\frac{d\Gamma'}{dt'})/\partial \Phi' & \partial(\frac{d\Gamma'}{dt'})/\partial \phi  & \partial(\frac{d\Gamma''}{dt'})/\partial \Gamma'  \end{array} \right] \]
evaluated at the equilibrium.
We write this symbolically as

\[M = \left[ \begin{array}{ccc}
\kappa \tilde{A}& \tilde{B}& 0 \\
\tilde{C}& \kappa \tilde{D}  & 1 \\
\kappa \tilde{E}& 0  &0  \end{array} \right] \]
where $\kappa$ denotes a term proportional to $1/ (n_{\Theta_1} \tau_a)$ coming from the non-conservative addition to the equations of motion and
\begin{align}
\kappa \tilde{A} & = -\frac{1}{\sqrt{2 \Phi'_{eq}}} \sin{\phi_{eq}}+ \frac{1}{Q|\beta| n_{\Theta_1}}C_0=\frac{C_0}{2Q|\beta| n_{\Theta_1}}  \nonumber \\
 \tilde{B} & = -\sqrt{2 \Phi'_{eq}} \cos{\phi_{eq}} \approx \sqrt{2 \Phi'_{eq}}+O(\kappa^2)\nonumber \\
\tilde{C} & = -1+\frac{1}{(2\Phi'_{eq})^{3/2}}\cos{\phi_{eq}}  \approx -\bigg(1+\frac{1}{(2\Phi'_{eq})^{3/2}}\bigg) + O(\kappa^2) \nonumber \\
\kappa \tilde{D} & = \frac{1}{\sqrt{2\Phi'_{eq}}}\sin{\phi_{eq}} =  \frac{C_0}{2Q|\beta| n_{\Theta_1}}\nonumber \\
\kappa \tilde{E} & = \frac{1}{Q|\beta| n_{\Theta_1}}(A_1+C_0) 
\end{align}
We will solve for the eigenvalues perturbatively in $\kappa$. At this stage,  we have taken advantage of the fact that the correction to $\cos{\phi_{eq}}=-1$ is second order in $\kappa$.

The equation for the eigenvalues is
\begin{align}
 0 =& (\kappa \tilde{A}-\lambda)(\kappa \tilde{D}-\lambda)(-\lambda)+\tilde{B}\kappa \tilde{E}+\lambda \tilde{B}\tilde{C}
\end{align}

We seek solutions of the form $\lambda = (\kappa \alpha_0,\kappa\alpha_1 \pm i (\alpha_2+\kappa \alpha_3))$.
Plugging in the first root yields an equation for $\alpha_0$. Plugging in the second and third roots yield 4 equations, for the real part and imaginary part of each root. Only two of these are distinct. Truncating at zeroth order in $\kappa$ yields an equation for $\alpha_2$, while at first order we find two equations for $\alpha_1$ and $\alpha_3$. We find that
\begin{align}
\alpha_0 & = -\frac{\tilde{E}}{\tilde{C}} \nonumber \\
\alpha_1 & = \frac{\tilde{A}+\tilde{D}}{2}+\frac{\tilde{E}}{2\tilde{C}} \nonumber \\
\alpha_2^2 & = -\tilde{B}\tilde{C}>0 \equiv \omega^2 \nonumber \\
\alpha_3 & = 0
\end{align}
where $\omega= \sqrt{|\tilde{B}\tilde{C}|}$ is the unperturbed frequency (of the conservative system) evaluated at the modified fixed point ($\Phi' = \Phi'_{eq}, \phi_{eq} \approx \pi)$.

We assume for simplicity that $m\approx (m+1), R\approx 1$ and $\alpha_{res} \approx 1$ (the compact approximation mentioned in the main text). Then:
\begin{align}\label{approx_compact}
\gamma_1 &=\frac{1}{\zeta+1} \nonumber \\
\eta_1 &=\bigg(\frac{1}{\zeta}+1 \bigg) \nonumber \\
 \gamma_2 &=\frac{\zeta}{\zeta+1} \nonumber \\
\eta_2 &= 1+\zeta \nonumber \\
Q & = \epsilon_p^{2/3} \frac{\zeta}{(1+\zeta)^2} \bigg(\frac{0.8^2 }{9m^2}\bigg)^{1/3} \nonumber \\
\Phi & =\frac{\zeta}{2(1+\zeta)}\sigma^2 \nonumber \\ 
\sigma^2 & = e_1^2+e_2^2-2e_1e_2\cos{\Delta \varpi}
\end{align}
 (where we have used Equation \eqref{eqn:phisigmarelation}), so that
\begin{align}\label{approxconstants}
C_0 & = -\frac{2}{\zeta+1}\frac{1}{\tau_{e}} +\frac{1-\zeta}{2(1+\zeta)}\frac{1}{\tau_{a}}\nonumber \\
A_0 & = \frac{\bigg(\frac{9}{0.8^2 m}\bigg)^{1/3}}{2 \epsilon_p^{2/3}  }\frac{1}{\tau_a}\nonumber \\
A_1 & = -\frac{2p}{m (1+\zeta)^2}\frac{1}{\tau_{a,e}} \nonumber \\
\end{align}

where
 \begin{align}\label{approxtimes}
  \frac{1}{\tau_{a} }& = \frac{1}{\tau_{a,2}}-\frac{1}{\tau_{a,1}} \nonumber \\
 \frac{1}{\tau_{e} }& = \frac{1}{\tau_{e,1}}+\frac{\zeta}{\tau_{e,2}} \nonumber \\
\frac{1}{ \tau_{a,e}} & = \frac{1}{\tau_{e,1}}-\frac{\zeta^2\alpha}{R^2\tau_{e,2}} \nonumber \\ 
 \end{align}
 Anticipating the discussion regarding $\tau_{a,e}$, we do not make the compact approximation {\it only} for this parameter.
 

 The equilibrium value of $\Phi'_{eq}$ can then be evaluated and used to determine the equilibrium value of $\sigma_{eq}$ by referring to Equation \eqref{approx_compact}, as
 \begin{align}\label{phieq_full}
\Phi'_{eq} &= -\frac{A_0}{A_1+C_0} = \frac{\bigg(\frac{9m^2}{0.8^2  \epsilon_p^2}\bigg)^{1/3}}{\frac{4m}{\zeta+1}\frac{\tau_a}{\tau_{e}}+\frac{4p}{ (1+\zeta)^2}\frac{\tau_a}{\tau_{a,e}}}= \sigma_{eq}^2\frac{1}{Q}\frac{\zeta}{2(1+\zeta)^2} \nonumber \\
\sigma_{eq}^2& =  \frac{1}{\frac{2m}{\zeta+1}\frac{\tau_a}{\tau_{e}}+\frac{2p}{ (1+\zeta)^2}\frac{\tau_a}{\tau_{a,e}}}\nonumber \\
\end{align}
where we have assumed that $\tau_a\gg\tau_{e}$ and hence ignored the second term in $C_0$. To obtain the equilibrium eccentricity without coupling between the semimajor axis and eccentricity evolution, set $p=0$. Also notice that if $\tau_{e,i} \rightarrow \infty$ (the limit of no eccentricity damping), both $\tau_{a,e}$ and $\tau_e$ diverge. In this case there is no equilibrium eccentricity.

The stability of the fixed point is determined by the sign of $\alpha_1$ and the sign of $\alpha_0$. First, the sign of $\alpha_1, s[\alpha_1]$, is given by:
 \begin{align}
 s[\alpha_1]& = s\bigg[ C_0(1+(2\Phi'_{eq})^{-3/2})-(A_1+C_0)\bigg] = s\bigg[ C_0(2\Phi'_{eq})^{-3/2}-A_1\bigg]\nonumber \\
C_0&  \approx -\frac{2}{\zeta+1}\frac{1}{\tau_{e}} \nonumber \\
A_1+C_0 & \approx-\frac{2p}{m (1+\zeta)^2}\frac{1}{\tau_{a,e}} -\frac{2}{\zeta+1}\frac{1}{\tau_{e}} 
 \end{align}
where we have assumed $\tau_e\ll\tau_a (K \gg 1)$. Then
 \begin{align}
  s[\alpha_1]& = s\bigg[ -\frac{2}{\zeta+1}\frac{1}{\tau_{e}} \bigg(\frac{2}{\zeta+1}\frac{\tau_a}{\tau_{e}}+\frac{2p}{m (1+\zeta)^2}\frac{\tau_a}{\tau_{a,e}}\bigg)^{3/2}\frac{0.8 \sqrt{m} \epsilon_p}{3}+\frac{2p}{m (1+\zeta)^2}\frac{1}{\tau_{a,e}} \bigg] \nonumber \\
  &= s\bigg[ -\bigg(\frac{2}{\zeta+1}\frac{\tau_a}{\tau_{e}}+\frac{2p}{m (1+\zeta)^2}\frac{\tau_a}{\tau_{a,e}}\bigg)^{3/2}\frac{0.8 \sqrt{m} \epsilon_p}{3}+\frac{p}{m (1+\zeta)}\frac{\tau_{e}}{\tau_{a,e}} \bigg] \nonumber \\
 \end{align}
Rearranging, $\alpha_1>0$ when:
 \begin{align}\label{critcrit}
\epsilon_p< \frac{3pm}{\mathcal{B}  2^{3/2}  }\frac{\tau_{e}}{\tau_{a,e}} \frac{(1+\zeta)^2}{\bigg(m(\zeta+1)+p\frac{\tau_{e}}{\tau_{a,e}}\bigg)^{3/2}} \bigg(\frac{\tau_{e}}{\tau_{a}}\bigg)^{3/2}
 \end{align}
 Here $\mathcal{B} = 0.8m$.  We remind the reader that the criterion given here has employed the compact approximation. If more accuracy is needed (likely only for the 2:1 resonance, refer to Figure \ref{fig:21}), return to the full expressions for $\tilde{A},\tilde{D},\tilde{C}$ and $\tilde{E}$, employing the full expressions for $\gamma_1,\gamma_2,\eta_1,\eta_2$ and $Q$, to evaluate $\alpha_1$ (or whatever else) directly.

 Importantly, the criterion Equation \eqref{critcrit} can never be satisfied if $\tau_{a,e}<0$, or if
 \begin{align}
\tau_{e,2}<\zeta^2 \alpha \tau_{e,1}/R^2 \nonumber \\
\tau_{e,2} \lesssim \zeta^2 \tau_{e,1}.
 \end{align}
(Again, the factors of $R$ and $\alpha$ are only important for the 2:1 commensurability).
 
In the case where $p=0$, 
 \begin{align}
   s[\alpha_1]& = s\bigg[ -\frac{2}{\zeta+1}\frac{1}{\tau_{e}} \bigg(\frac{2}{\zeta+1}\frac{\tau_a}{\tau_{e}}\bigg)^{3/2}\frac{0.8 \sqrt{m} \epsilon_p}{3} \bigg]  = s\bigg[ -\frac{1}{\tau_{e}} \bigg(\frac{\tau_a}{\tau_{e}}\bigg)^{3/2} \bigg] 
 \end{align}
  which is always negative for convergent migration $\tau_a>0$ and eccentricity damping $\tau_e>0$. This shows the stability of the fixed point when the eccentricity-semimajor axis coupling term is not present.

Next, we turn to the sign of $\alpha_0$. Since $\tilde{C}<0$,  $s[\alpha_0]$  is determined by $s[\tilde{E}]$:
\begin{align}
s[\alpha_0]& = s[\tilde{E}] = s[A_1+C_0]  \approx s[-\frac{p}{m (1+\zeta)}\frac{\tau_{e}}{\tau_{a,e}} -1 \bigg]
\end{align} 

Note, however, that for $\Phi'_{eq}$ to be a positive quantity (for the fixed point to exist),  its denominator must be positive. From the definition given in Equation \eqref{phieq_full}, we see that this means that (canceling common positive factors between the two terms in the denominator)  $1+(p/m)(\tau_e/\tau_{a,e})(1+\zeta)^{-1}>0$. In that case, $s[\alpha_0]<0$ always.

\subsection{Limiting case of a test particle}
\subsubsection{Inner test particle}

We consider the case where the inner planet is a test particle, moving outwards, towards a massive planet on a fixed circular orbit. Then $\zeta \rightarrow 0,\tau_{a,1}\rightarrow -\tau_{a,1}, \tau_{a,2}\rightarrow \infty$, $\tau_{e,2} \rightarrow \infty$ and
\begin{align}
\frac{1}{\tau_a} & = \frac{1}{\tau_{a,1}} \nonumber \\
\frac{1}{\tau_{a,e}} & =\frac{1}{\tau_{e}} =  \frac{1}{\tau_{e,1}} 
\end{align}
The criterion is 
 \begin{align}
\epsilon_2< \frac{3pm}{\mathcal{B}  2^{3/2}  } \frac{1}{(m+p)^{3/2}} \bigg(\frac{\tau_{e,1}}{\tau_{a,1}}\bigg)^{3/2}\nonumber \\
\end{align}
Plugging in $p=1$ and converting from $\tau_{a,1}$ to $\tau_{n,1}$ yields:
\begin{align}
\epsilon_2< \bigg(\frac{m}{\mathcal{B} \sqrt{3}(1+m)^{3/2}}\bigg) \bigg(\frac{\tau_{e,1}}{\tau_{n,1}}\bigg)^{3/2}  
\end{align}
The equilibrium value for eccentricity is given by:
\begin{align}
e_1^2 & = \frac{1}{2(p+m)}\frac{\tau_{e,1}}{\tau_{a,1}} =  \frac{1}{3(p+m)}\frac{\tau_{e,1}}{\tau_{n,1}}\bigg\lvert_{p=1} =  \frac{1}{3(1+m)}\frac{\tau_{e,1}}{\tau_{n,1}}
\end{align}
The equilibrium values of $\sin{\phi}$ is given as:
\begin{align}
\sin{\phi_{eq}} &= \frac{\sqrt{2\Phi'_{eq}}}{2 Q|\beta| n_{\Theta_1}} C_0 \nonumber \\
Q |\beta| n_{\Theta_1} &= n_2 \epsilon_p^{2/3}(0.8^2m^4 3)^{1/3} \nonumber \\
\sin{\phi_{eq}}& = -\frac{e_1}{ \tau_{e,1} n_1  \epsilon_p \mathcal{B}} \nonumber \\
\end{align}
which was found by \citet{Sch}.
\subsubsection{Outer test particle}
In this case, $\tau_{a,1}\rightarrow \infty, \tau_{e,1} \rightarrow \infty,$ and $\zeta \rightarrow \infty$. The equilibrium eccentricity is given by
\begin{align}
e_2^2 = \frac{\frac{\tau_{e,2}}{\tau_{a_2}}}{2(m-p)} = \frac{1}{2K(m-p)}
\end{align}
We emphasize that this estimate of the equilibrium eccentricity is an approximation. It does appear that for some parameter choices ($m\geq p$), the equilibrium eccentricity does not exist. This means that there is no fixed point in resonance. According to \citet{Tanaka2}, $p<1$, so since $m\geq 1$ this issue may not arise in practice. However, in the case of $p=1$, we have confirmed that the full expression for the equilibrium eccentricity does not yield a diverging equilibrium for $m=1$ in the case of the outer test particle (see Figure \ref{fig:testPart}). 

The criterion for over stability is
 \begin{align}
\epsilon_1<  -\frac{3pm}{\mathcal{B}  2^{3/2}} \frac{1}{(m-p)^{3/2}} \bigg(\frac{\tau_{e,2}}{\tau_{a,2}}\bigg)^{3/2}
 \end{align}

This implies that $\alpha_1$ is never positive.  Note that we assume $m>p$ to get the equilibrium, which also implies that $\alpha_0<0$, and that the criterion for over stability $\epsilon_{1,crit}$ is a real number.

\subsection{The assumption that $\Psi_2=0$}\label{sec:psi2zero}
 Since we have prescribed eccentricity damping, the assumption that $\Psi_2$ is zero when the resonance is reached is a good one. However, if the system is captured into resonance, the eccentricities grow as $\Phi$ grows, and the condition that $\Psi_2=0$ requires that this growth preserves the orientation of the orbits and the ratio of the eccentricities. This is true because when $\Psi_2=0$, $r_2=0$ and $s_2=0$, which in turn imply that 
\begin{align}\label{important}
 \varpi_1-\varpi_2  & = \pi \nonumber \\
 P_1 & = \bigg(\frac{\bar{\delta}_1}{\bar{\delta}_2}\bigg)^2 P_2 \nonumber \\
\mbox{or} \nonumber \\
e_2 & = e_1 \zeta \sqrt{\alpha}/R \nonumber \\
\tan{\psi} & = \tan{p_1} 
\end{align} 
with $R = |f_{27}/(f_{31}-2\alpha\delta_{m,1})|$. 

It cannot be true that $\Psi_2=0$ generically for large eccentricities. For example, it has been demonstrated numerically and analytically that for the 2:1 resonance the center of resonance changes as the eccentricities grow, from anti-alignment of the orbits ($\varpi_1 = \varpi_2+\pi$) to exact alignment ($\varpi_1 = \varpi_2$) or asymmetric libration, depending on the mass ratio of the planets to each other \citep{LeePeale,Beauge}. This change implies $\Psi_2$ deviates from zero.

To test how large the eccentricities can be before the approximation breaks down, we performed numerical tests, driving systems with $\Psi_2(t=0) =0$ into resonance without eccentricity damping. This leads to an unchecked increase in $\Phi'$ and in $\sigma \approx \sqrt{e_1^2+e_2^2-2e_1e_2\cos{\Delta \varpi}}$. If this growth preserves $\Psi_2=0$, the eccentricities will grow along a track of constant $e_1/e_2$. Eccentricity damping, if included, would halt the system at a particular value of $\sigma$ along this track. 

\begin{figure}
	\begin{center}
	\includegraphics[width=3.5in]{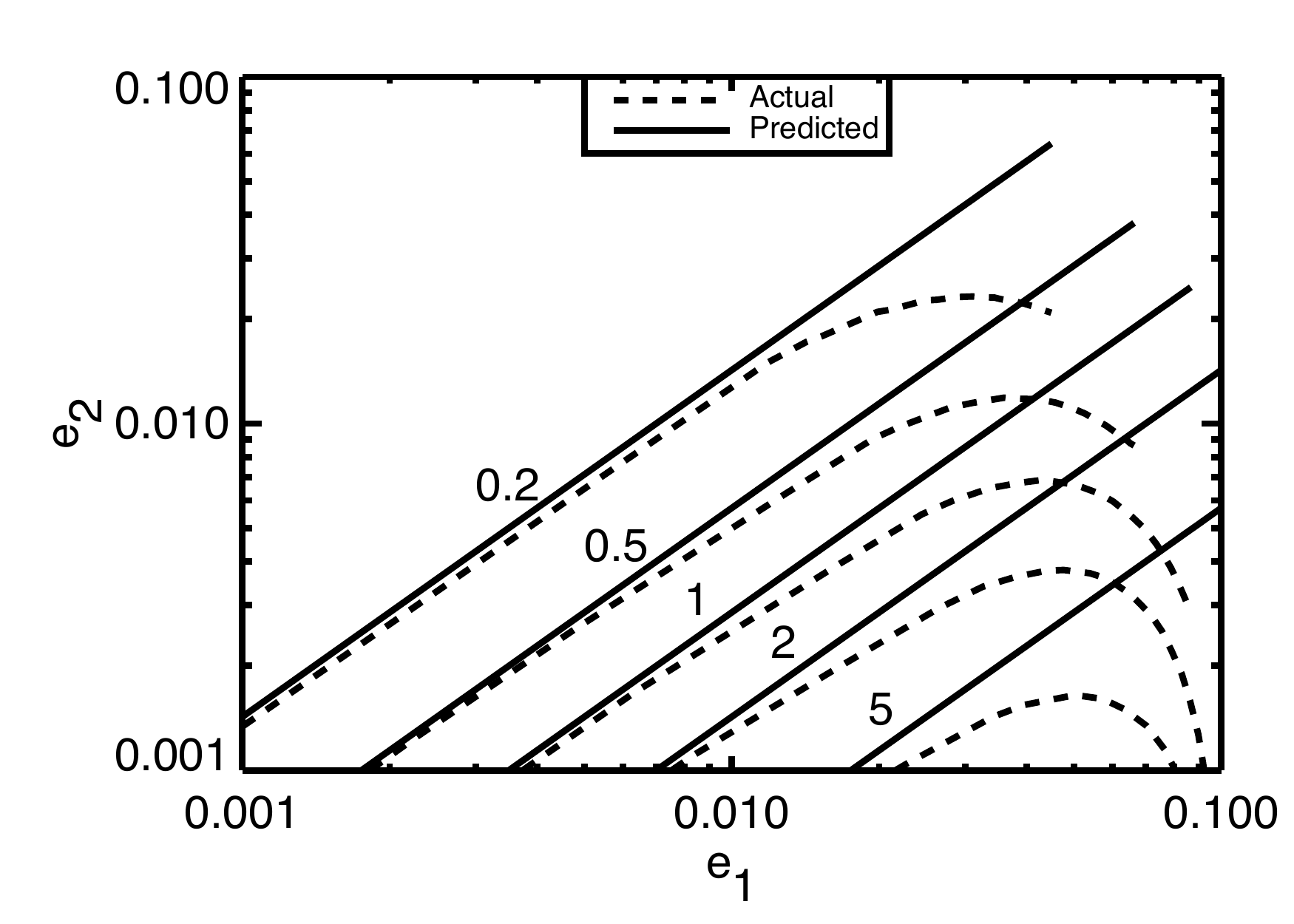} 
	\caption{Comparison of the predicted equilibrium eccentricities in the case of $\Psi_2=0$ (solid) with observed equilibrium eccentricities (dashed) for the 2:1 mean motion resonance for different values of the planetary mass ratio $\zeta$ (as labeled). The magnitude of the eccentricity damping determines where the system halts along the dashed curve. }
	\label{fig:twoto1}
	\end{center}
	\end{figure}

\begin{figure}
	\begin{center}
	\includegraphics[width=3.5in]{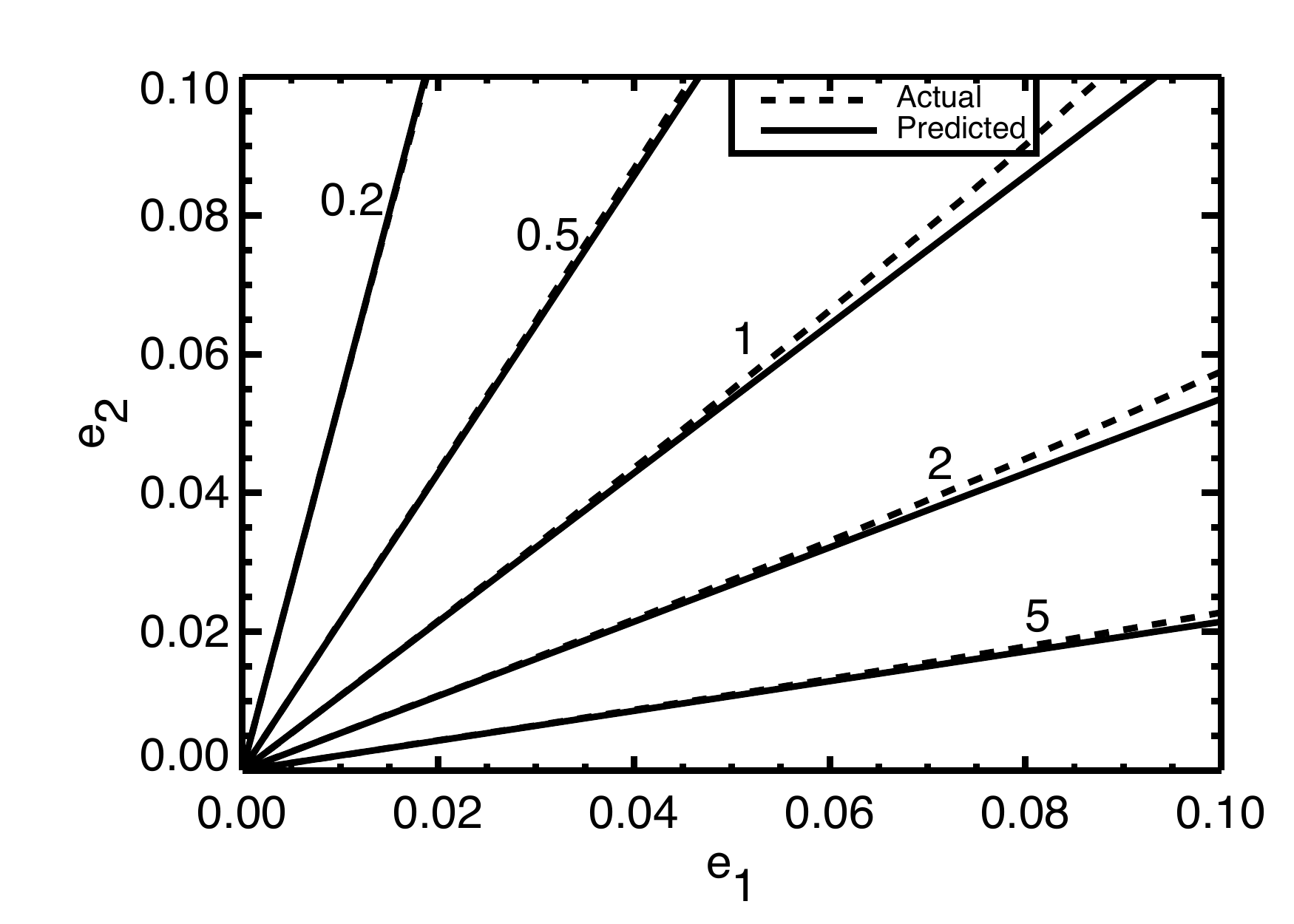}
	\caption{Comparison of the predicted equilibrium eccentricities in the case of $\Psi_2=0$ (solid) with observed equilibrium eccentricities (dashed) for the 3:2 mean motion resonance for different values of the planetary mass ratio $\zeta$ (as labeled). The magnitude of the eccentricity damping determines where the system halts along the dashed curve.  }
	\label{fig:threeto2}
	\end{center}
	\end{figure}

 In Figure \ref{fig:twoto1}, we show, as a function of $\zeta$, the comparison between the analytic prediction with $\Psi_2 =0$ and the results of the numerical integrations for $e_1$ and $e_2$ as the system is driven into the 2:1 mean motion resonance. We show the same for the 3:2 resonance in Figure \ref{fig:threeto2}. How low eccentricities need to be (or equivalently, how effective the damping must be) for the assumption that $\Psi_2\approx 0$ to be good is also a function of the resonance integer $m$. Especially for the 2:1 resonance, it is clear that the true evolution drives $\Psi_2$ away from zero as eccentricities grow, but for the 3:2 resonance the prediction based on $\Psi_2=0$ is very good across the entire range studied $(e_1,e_2)\in [0,0.1]$. For the 2:1 resonance in particular, depending on the equilibrium eccentricities, the predictions of the theory may be only roughly correct, and they may be wrong entirely in the case where the resonance center corresponds to alignment or asymmetric alignment of the orbits. 

\end{document}